# RADIATIVE NEUTRON CAPTURE ON $^9$Be, $^{14}$C, $^{14}$N, $^{15}$N AND $^{16}$O AT THERMAL AND ASTROPHYSICAL ENERGIES

SERGEY DUBOVICHENKO[1,2,*], ALBERT DZHAZAIROV-KAKHRAMANOV[1,2,†] and NADEZHDA AFANASYEVA[1,3,‡]

[1]*V. G. Fessenkov Astrophysical Institute "NCSRT" NSA RK, 050020, Observatory 23, Kamenskoe plato, Almaty, Kazakhstan*
[2]*Institute of nuclear physics NNC RK, 050032, str. Ibragimova 1, Almaty, Kazakhstan*
[3]*Al-Farabi Kazakh National University, 050040, av. Al-Farabi 71, Almaty, Kazakhstan*
[*]*dubovichenko@mail.ru*
[†]*albert-j@yandex.ru*
[‡]*n.v.afanasyeva@gmail.com*

The total cross sections of the radiative neutron capture processes on $^9$Be, $^{14}$C, $^{14}$N, $^{15}$N, and $^{16}$O are described in the framework of the modified potential cluster model with the classification of orbital states according to Young tableaux. The continued interest in the study of these reactions is due, on the one hand, to the important role played by this process in the analysis of many fundamental properties of nuclei and nuclear reactions, and, on the other hand, to the wide use of the capture cross section data in the various applications of nuclear physics and nuclear astrophysics, and, also, to the importance of the analysis of primordial nucleosynthesis in the Universe. This article is devoted to the description of results for the processes of the radiative neutron capture on certain light atomic nuclei at thermal and astrophysical energies. The considered capture reactions are not part of stellar thermonuclear cycles, but involve in the reaction chains of inhomogeneous Big Bang models.

## 1. Introduction

This review is the logical continuation of works devoted to the radiative proton and neutron capture on light nuclei that were published in the International Journal of Modern Physics E **21** (2012) 1250039-1 and **22** (2013) 1350028-1.

### 1.1. *Nuclear aspects of the review*

The extremely successful development of microscopic models like resonating group method (RGM), see, for example, Refs. 1 and 2, generator coordinate method (GCM), see, particularly, Ref. 3 or algebraic version of RGM,[4] leads to the view that the advancement of new results in low energy nuclear physics and nuclear astrophysics is possible only in this direction. Eventually, very common opinion states that this is the only way in which the future development of our ideas about structure of atomic nucleus, nuclear and thermonuclear reactions at low and astrophysical energies can be imagined.

However, the possibilities of simple two-body potential cluster model (PCM) were not completely studied up to now, particularly, if it use the conception of forbidden states (FSs)[5] and directly take into account the resonance behavior of elastic scattering phase shifts of interacting particles at low energies.[6] This model can be named as the modified PCM (MPCM with FS or simply MPCM). The potentials of the bound states (BSs) in this model are constructed on the basis of the description not only the phase shift, binding energy and charge radii, but also the asymptotic constants (AC) in the specified channel. The rather difficult RGM calculations are not the only way to explain the available experimental facts. Simpler the MPCM with FS can be used, taking into account the classification of orbital states according to Young tableaux and the resonance behavior of the elastic scattering phase shifts. Such approach, as it was shown earlier,[5,6,7] in many cases allows one to obtain quite adequate results in description of many experimental data.

The results of the phase shift analysis of the experimental data on the differential cross sections for the elastic scattering[8] of corresponding free nuclei correlated to such clusters[9,10,11,12] are usually used for the construction of the interaction potentials between the clusters for the scattering states in the MPCM. Intercluster interaction potentials within the framework of the two-particle scattering problem are constructed from the best description of the elastic scattering phases.[12,13,14] Moreover, for any nucleon system, the many-body problem is taken into account by the division of single-particle levels of such potential on the allowed and the forbidden by the Pauli principle states.[9-12] However, the results of phase shift analysis, being usually available only in a limited range of energies and with large errors, do not allow, as a rule, to reconstruct completely uniquely the interaction potential of the scattering processes and the BSs of clusters. Therefore, an additional condition for the construction of the BS potential is a requirement of reproduction of the nucleus binding energy in the corresponding cluster channel and a description of some other static nuclear characteristics, such as the charge radius and the AC.

This additional requirement is, of course, an idealization of real situations in the nucleus, because it suggests that there is 100% clusterization in the ground state. Therefore, the success of this potential model to describe the system of $A$ nucleons in a bound state is determined by how high the real clusterization of this nucleus in the $A_1 + A_2$ nucleons channel. At the same time, some of the nuclear characteristics of the individual, not even the cluster, nuclei can be mainly caused by one particular cluster channel. In this case, the single-channel cluster model used here allows us to identify the dominant cluster channel, to identify and describe those properties of nuclear systems that are caused by it.[9-11] In addition, the potentials of the BS and the continuous spectrum, taking into account the resonant nature of the elastic scattering phase shifts in a given partial wave must meet the classification of cluster orbital states according to the Young tableaux.

## 1.2. *Astrophysical aspects of the review*

As we know, light radioactive nuclei play an important role in many astrophysical environments. In addition, such parameter as cross section of the capture reactions as a function of energy is very important for investigation of many astrophysical problems such as primordial nucleosynthesis of the Universe, main trends of stellar evolution, novae and super-novae explosions, X-ray bursts etc. The continued interest in the study of processes of radiative neutron capture on light nuclei at low and ultralow energy is caused by several reasons. Firstly, this process plays a significant part in the study of many fundamental properties of nuclear reactions, and secondly, the data on the capture cross sections are widely used in a various applications of nuclear physics and nuclear astrophysics, for example, in the process of studying of the primordial nucleosynthesis reactions.

In the beginning of this work we will consider the possibility to describe the total cross sections of the neutron capture on $^9$Be on the basis of the MPCM, where the supermultiplet symmetry of wave function and the separation of orbital states according to Young tableaux are taken into account.

The reaction $^9$Be(n, $\gamma$)$^{10}$Be is important in the $r$-process of the Type-II supernovae.[15] In the early expanding phase of the Type-II supernovae, the mechanism to bypass the stability gap is going through processes, $\alpha + \alpha + \alpha \rightarrow {}^{12}$C or $\alpha + \alpha + n \rightarrow {}^9$Be. The $r$-process is triggered after neutron-rich freeze-out. So, sufficient neutron abundance, the $A = 8$ mass gap could be alternatively found away through the Be-isotope chain:

$$^4\text{He}(\alpha n, \gamma)^9\text{Be}(n, \gamma)^{10}\text{Be}(\alpha, \gamma)^{14}\text{C} \tag{1}$$

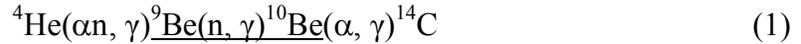

or through the Li-B chain.[15,16] Therefore, the reactions rates of $^9$Be(n, $\gamma$)$^{10}$Be would influence the abundance of the heavier isotopes in the Type-II supernovae.

Furthermore, we will study the $^{14}$C(n, $\gamma$)$^{15}$C reaction. Therefore, continuing the analysis of the radiative capture of neutrons by light atomic nuclei, which is part of the different thermonuclear cycles[17,18,19,20] and take part in one of the variants of reaction chains for inhomogeneous Big Bang models[21,22,23,24,25] this allow to the synthesis of heavy elements via the next chain of neutron capture reactions

$$^1\text{H}(n,\gamma)^2\text{H}(n,\gamma)^3\text{H}(^2\text{H},n)^4\text{He}(^3\text{H},\gamma)^7\text{Li}(n,\gamma)^8\text{Li}(^4\text{He},n)^{11}\text{B}(n,\gamma)^{12}\text{B}(\beta^-)$$
$$^{12}\text{C}(n,\gamma)^{13}\text{C}(n,\gamma)^{14}\text{C}(n,\gamma)^{15}\text{C}(\beta^-)^{15}\text{N}(n,\gamma)^{16}\text{N}(\beta^-)^{16}\text{O}(n,\gamma)^{17}\text{O}(n,^4\text{He})^{14}\text{C}\ldots \tag{2}$$

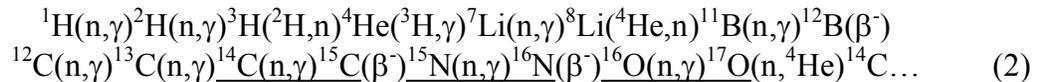

The half life of $^{15}$C is equal to $T_{1/2} = 2.449(5)$ sec. Here, we will dwell on this radiative capture process at energies 23 keV–1.0 MeV. The $^{14}$C(n, $\gamma$)$^{15}$C reaction is also a part of the neutron induced CNO cycles in the helium burning layer of Asymptotic Giant Branch (AGB) stars, in the helium burning core of massive stars, and in subsequent carbon burning.[26] Such cycles may cause a depletion in the CNO abundances. The $^{14}$C(n, $\gamma$)$^{15}$C reaction is the slowest of both of these cycles and, therefore the knowledge of its rate is important to predict the $^{14}$C abundances.[27]

Then the n$^{14}$N $\rightarrow$ $^{15}$N$\gamma$ capture reaction, which leads to the formation and accumulation of the $^{15}$N nuclei, will be considered. Thereby, this reaction is additional to the $^{14}$C(n,$\gamma$)$^{15}$C($\beta^-$)$^{15}$N process, which take place in the main reaction chain and increases the amount of $^{15}$N, taken part in subsequent synthesis reactions of heavier elements. Another interesting effect of bombarding $^{14}$N by neutrons is the following reaction: n$^{14}$N $\rightarrow$ $^{14}$Cp. Carbon-14 is produced in the upper layers of the troposphere and the stratosphere by thermal neutrons, which are the result of interaction between cosmic rays and atmosphere. The highest rate of carbon-14 production takes place at altitudes of 9 to 15 km and at high geomagnetic latitudes. The reaction n$^{14}$N $\rightarrow$ $^{14}$Cp is the main source of isotope $^{14}$C in the Earth (99.634%).[28]

Furthermore, we will consider the radiative capture reaction $^{15}$N(n, γ)$^{16}$N. This neutron capture reaction on light nuclei may be of considerable importance for the *s-process* (slow-neutron-capture-process) nucleosynthesis in red giant stars as well as for the nucleosynthesis in inhomogeneous Big Bang scenarios (see Eq. (2)). To determine the reaction rates for such different temperature conditions, the cross sections of radiative capture need to be obtained for a wide energy range.

Finally, we will consider the reactions of neutron capture on $^{16}$O at thermal and astrophysical energies for analysis of the processes of radiative capture of neutrons by light atomic nuclei. This reaction is also important for *s*-processes for stars of various metallicity.[6,7,17,18] It also considered in the frame of inhomogeneous Big Bang models[21-25] which, for masses beyond $A > 12$ can proceed via Eq. (2) (…$^{15}$N(n,γ)$^{16}$N(β$^-$) $^{16}$O(n,γ)$^{17}$O(n,$^4$He)$^{14}$C…).

## 2. Model and calculation methods

The expressions for the total radiative capture cross-sections σ($NJ,J_f$) in the potential cluster model are given, for example, in Refs. 29 and 30 are written as

$$\sigma_c(NJ,J_f) = \frac{8\pi Ke^2}{\hbar^2 q^3} \frac{\mu}{(2S_1+1)(2S_2+1)} \frac{J+1}{J[(2J+1)!!]^2}$$
$$\times A_J^2(NJ,K) \sum_{L_i,J_i} P_J^2(NJ,J_f,J_i) I_J^2(J_f,J_i) \quad (3)$$

where for the electric orbital $EJ(L)$ transitions ($S_i = S_f = S$) we have:[29]

$$P_J^2(EJ,J_f,J_i) = \delta_{S_iS_f}\left[(2J+1)(2L_i+1)(2J_i+1)(2J_f+1)\right](L_i 0 J 0 | L_f 0)^2 \begin{Bmatrix} L_i & S & J_i \\ J_f & J & L_f \end{Bmatrix}^2,$$

$$A_J(EJ,K) = K^J \mu^J \left(\frac{Z_1}{m_1^J} + (-1)^J \frac{Z_2}{m_2^J}\right), \quad I_J(J_f,J_i) = \langle \chi_f | R^J | \chi_i \rangle. \quad (4)$$

Here, $q$ is the wave number of the initial channel particles; $L_f$, $L_i$, $J_f$, $J_i$, $S_f$, and $S_i$ are the angular moments of particles in the initial (*i*) and final (*f*) channels; $S_1$ and $S_2$ are spins of the initial channel particles; $m_1$, $m_2$, $Z_1$, and $Z_2$ are the masses and charges of the particles in the initial channel, respectively; $K$ and $J$ are the wave number and angular moment of the γ-quantum in the final channel; and $I_J$ is the integral over wave functions of the initial $\chi_i$ and final $\chi_f$ states as functions of relative cluster motion with the intercluster distance $R$. Emphasize that in our calculations here or earlier, we have never used such a notion as a spectroscopic factor (see, for example, Ref. 29), i.e., its value is simply assumed to be equal to 1.[9-12,30]

For the spin part of the $M1(S)$ magnetic process in the used model the following expression is known ($S_i = S_f = S$, $L_i = L_f = L$)[29,30]

$$P_1^2(M1,J_f,J_i) = \delta_{S_iS_f} \delta_{L_iL_f}\left[S(S+1)(2S+1)(2J_i+1)(2J_f+1)\right] \begin{Bmatrix} S & L & J_i \\ J_f & 1 & S \end{Bmatrix}^2,$$

$$A_1(M1,K) = \frac{e\hbar}{m_0 c} K\sqrt{3}\left[\mu_1 \frac{m_2}{m} - \mu_2 \frac{m_1}{m}\right], \quad I_J(J_f, J_i) = \langle \chi_f | R^{J-1} | \chi_i \rangle, \quad J = 1. \tag{5}$$

Here, $m$ is mass of the nucleus, $\mu_1$, $\mu_2$ are the magnetic moments of clusters and the remaining symbols are the same as in the previous expression. The value of $\mu_n$ = -1.91304272$\mu_0$ is used for the magnetic moment of neutron.[31] The correctness of the given above expression for the $M1$ transition is pre-tested on the basis of the radiative proton and neutron captures on $^7$Li and $^2$H at low energies in our previous studies.[6,7,32,33,34]

To perform the calculations of the total cross section our computer program based on the finite-difference method (FDM)[35] has been rewritten. Now the absolute search precision of the binding energy for n$A$ system for the different potentials is equal to $10^{-6}$–$10^{-8}$ MeV, the search precision of the determinant zero in the FDM, which determines the accuracy of the binding energy search, is equal to $10^{-14}$–$10^{-15}$, and the magnitude of the Wronskians of the Coulomb scattering wave functions is about $10^{-15}$–$10^{-20}$.[35]

The variational method (VM) with the expansion of the cluster wave function (WF) of the relative motion of the considered systems on a non-orthogonal Gaussian basis[12,35] was used for additional control of the binding energy calculations:[12,35,36,37,38]

$$\Phi_L(r) = \frac{\chi_L(r)}{r} = Nr^L \sum_i C_i \exp(-\beta_i r^2), \tag{6}$$

where $\beta_i$ and $C_i$ are the variational parameters and expansion coefficients, $N$ is the normalization coefficient of WF.

A computer program for solving of the generalized variational problem[35] has also been modified and used to control the accuracy of the FDM calculating of the binding energy and the wave function form. One of the possible and simple enough numerical algorithms for solving this problem with a simple program realization has been considered in Refs. 35 and 39.

The asymptotic constant of the ground state potential normally used to control the behavior of the wave function of the bound state at large distances, was calculated using the asymptotic behavior of the wave function in the Whittaker form[40]

$$\chi_L(r) = \sqrt{2k} C_W W_{-\eta_L + 1/2}(2kr), \tag{7}$$

where $\chi$ is the numerical bound state wave function obtained from the solving of the Schrödinger equation and normalized to unity, $W$ is the Whittaker function of the bound state, which determines the asymptotic behavior of the wave function and it is a solution of the same equation without the nuclear potential, i.e. the solution at large distances, $k$ is the wave number caused by the binding energy of the channel, $\eta$ is the Coulomb parameter defined hereinafter, $L$ is the orbital angular momentum of the bound state.

The intercluster interaction potentials for considered systems, as usual, are chosen in simple Gaussian form:[5,7]

$$V(r) = -V_0 \exp(-\alpha r^2), \tag{8}$$

where $V_0$ and $\alpha$ are the potential parameters usually obtained on the basis of description of the elastic scattering phase shifts at certain partial waves taking into account their resonance behavior or spectrum structure of resonance levels and, in the case of discrete spectrum, on the basis of description of the BS characteristics of the n$^9$Be system. In both cases such potentials contain BSs, which satisfy the classification of allowed states (ASs) and FSs according to Young tableaux given above.

In the present calculations the exact values of the neutron mass $m_n$ = 1.00866491597 amu,[41] the mass of $^9$Be is equal to 9.012182 amu, the mass of $^{14}$C is equal to 14.003242 amu, the mass of $^{14}$N is equal to 14.003074 amu, the mass of $^{15}$N is equal to 15.000108 amu, and the mass of $^{16}$O equals 15.994915 amu.[42] The value of the $\hbar^2/m_0$ constant is equal to 41.4686 MeV fm$^2$. The Coulomb parameter $\eta = \mu Z_1 Z_2 e^2/(k\hbar^2)$ equals zero in this case, is represented as $\eta = 3.44476 \cdot 10^{-2} \cdot Z_1 \cdot Z_2 \cdot \mu/k$, where $k$ is the wave number in the fm$^{-1}$, defined by the interaction energy of the particles and $\mu$ is the reduced mass of the particles in amu.

## 3. Radiative neutron capture on $^9$Be in cluster model

First, we note that in the framework of the approach using here, notably, the MPCM with the classification of cluster states according to Young tableaux, with states forbidden by the Pauli principle. Earlier, there were considered 20 capture reactions of protons, neutrons and light clusters on 1$p$-shell nuclei at astrophysical energies by this method.[6,7,11,43,44] Therefore, we will consider the n$^9$Be → $^{10}$Be$\gamma$ reaction at thermal and astrophysical energy range in the study of the neutron radiative capture reactions on light atomic nuclei in the MPCM.[5,7,30] The information about n$^9$Be interaction potentials in continuous and discrete spectra[7,45] is necessary for calculating the total cross-sections of this reaction in the frame of the MPCM.[5,7] As before, we think that these potentials should correspond to the classification of the cluster state according to orbital symmetries,[5] as in our works[11,33,43,44] for other nuclear systems taking part in different thermonuclear processes or reactions of primordial nucleosynthesis.[17]

### 3.1. *Classification of the orbital states*

First, we draw attention to the classification of the cluster orbital states on the basis of Young tableaux for $^9$Be. If it is possible to use tableaux {44} and {1} in the 8 + 1 particle system, then in the issue of exterior product {44} × {1}, two possible orbital symmetries, {54} + {441}, will be obtained for $^9$Be. The first is forbidden as it contains five cells in one row, and the second is ranked as the allowed Young tableaux, which corresponds to the allowed state of the cluster relative motion of n$^8$Be in $^9$Be.[46] Let us note that the classification of orbital states according to Young tableaux given here has only qualitative character, because for systems with $A$ = 9 and 10 we cannot find tables of inner products for Young tableaux, which determine the spin-isospin symmetry of the cluster system WF. These data were available for all $A < 9$ (see Ref. 47) and were previously used by us for the analysis of the number of ASs and FSs in wave functions for different cluster systems.[7,9,30]

Furthermore, the tableau {441} corresponds to the ground state (GS) of $^9$Be, so the N$^9$Be system contains the FS with tableau {541} for $L = 1$ and 2 and the AS with configuration {4411} with $L = 1$. Here we are limited to the minimal values of orbital moments $L$, which are required in future. The bound state in the $S$ and $D$ waves with the tableau {442} is also present. Let us note that because of the absence of tables of internal products of the Young tableaux for $A > 8$, it is impossible to unambiguously maintain – is this state forbidden or allowed? Therefore, to fix the idea we will consider that there are the bound AS in the $^3S$ wave with {442}, and in the $^3D$ wave, which contains the bound FS for tableau {541}; the tableau {442} is unbound. Since, the GS of $^{10}$Be with $J = 0^+$ can be formed only in the triplet spin state $^3P_0$, then further for all BSs in the n$^9$Be channel we accept $S = 1$. Though, actually, some of the bound and scattering states could be mixed according to spin with $S = 1$ and 2.

Thereby, N$^9$Be potentials in the $^3P_0$ wave should have forbidden and allowed bound states with Young tableaux {541} and {4411}, first of them is forbidden and the second is allowed – it corresponds to the GS of $^{10}$Be in the n$^9$Be channel. We will consider the AS for {4411} in the other $^3P$ waves as unbound, i.e., their potentials will contain only one bound FS with tableau {541}. The potential, in which we match the resonance state of nucleus at the energy of 5.9599 MeV for $J^\pi = 1^-$ relatively to the GS of $^{10}$Be and bound in the n$^9$Be channel, was considered as a variant for the potential of the $^3S_1$ scattering wave with one BS for tableau {442}. Thus, we unambiguously fix the structure of FSs and ASs in each partial potential for $L = 0, 1, 2$, which will be considered. The similar situation was observed earlier in Ref. 44 for n$^{12}$C and n$^{13}$C systems in the radiative neutron capture processes. Note that the number of BSs, FSs or ASs, in any partial potential, determines the number of WF nodes at low distances, usually less than 1 fm.$^5$ It will be recalled that the WF of the BS with the minimal energy does not have nodes; the next BS has one node etc.

In our previous work of Ref. 48, on the basis of consideration of the transitions only to the GS of $^{10}$Be from the $^3S$ scattering wave with the zero phase shift, it was shown that the considered ambiguity of the number of forbidden or allowed BSs in the $^3S_1$ scattering wave and the $^3P_0$ potential of the GS of $^{10}$Be practically does not influence on the calculation results, if these potentials contains 1-2 BSs in the first case or from 2 to 3 BSs in the second. In the first case, the state with the maximum energy corresponds to the tableau {442} and is allowed, in the second case, such state is allowed also and corresponded to the GS of $^{10}$Be in the n$^9$Be channel with the tableau {4411}. Therefore, the potentials of the $^3S$ and $^3P_0$ waves can be matched with the given above classification of FSs and ASs according to Young tableaux, considered in Ref. 48 as the second variant.

Furthermore, we will consider transitions from the $^3D$ scattering wave, and as a ground potential we accept the variant with one FS with tableau {541}, considering that the unbound AS corresponds to tableau {442}. The results of calculations with this potential are compared with two other its variants. At first place, we are considering the variant of the potentials without BSs at all, which is not agree with the given above classification. As the second option, the potential with two BSs, first of them is forbidden with tableau {541}, and the second is allowed and bound with {442}. Such variant is also matched with the given above classification according Young tableaux, because it does not give an opportunity to identify – is the AS bound or not?

It is necessary to keep in mind that for the *S* wave the internal part of the nucleus is "transparent" due to absence of the Coulomb and angular momentum barrier at the neutron capture and the number of nodes in WF at low distances plays appreciable role – results for WF without nodes and with one-two nodes differ from each other noticeably.[48] The angular momentum barrier exists in the case of the *D* wave, therefore the dependence of results from the structure of the *D* wave in the internal range, i.e., from the number of nodes at low distances, will be noticeably weaker than in the previous case.

### 3.2. *Potential description of the scattering phase shifts*

Now, we would like to present more details on the construction procedure of the intercluster potentials used here, defining the criteria for finding these parameters and the order in which they are found. Primarily, the parameters of the GS potential, which are determined by the given number of allowed and forbidden states in this partial wave, are fixed quite unambiguously by the binding energy, the charge radius of the nucleus, and the asymptotic constant. The accuracy of determination of the GS potential, in the first place, is connected with the AC accuracy, which is usually equal to 10–20%.[7] There are no other ambiguities in this potential, because the classification of the states according to Young tableaux allows us unambiguously fix the number of BS, which completely determines its depth and the width of the potential fully depends on the AC value. Consequently, we obtain completely unambiguous potential with errors of parameters, determined by the spread of AC values.

The intercluster potential of the non-resonance scattering process for the case of the $^3S_1$ wave, constructed according to the scattering phase shifts at the given number of allowed and forbidden BSs in the considered partial wave, is also fixed quite unambiguously. The accuracy of determination of this potential is connected, in the first place, with the accuracy of the derivation of the scattering phase shift from the experimental data and is usually equal to about 20–30%. It is difficult to estimate the accuracy with which parameters of the scattering potential are found during its construction according to the nuclear spectrum data in certain channels even at the given number of BSs, although apparently it can be hoped that the error will not be much bigger than in the previous case.

We have not had any success in the data search for the n$^9$Be elastic scattering phase shifts at astrophysical energies,[49,50,51] and therefore the $^3S_1$ potential of the scattering process that leads to zero scattering phase shifts at energies up to 1 MeV will be considered here. This follows from the data of the energy level spectrum of $^{10}$Be, which does not contain *S*-resonances with $J^\pi = 1^-$ in this energy range.[52,53]

The potential is constructed completely unambiguously with the given number of BSs and with the analysis of the resonance scattering when in the considered partial wave at the energies up to 1 MeV there is a rather narrow resonance with a width of about 10–50 keV. The error of its parameters does not usually exceed the error of the width determination at this level and equals 3–5%. The depth of the potential at the given number of BSs completely depends on the resonance energy, and its width is determined by the width of this resonance.

Now let us go to the direct construction of the intercluster interactions – the potential of the $^3P_0^1$ ground bound state is constructed on the basis of description of the nuclear characteristics of $^{10}$Be in the n$^9$Be system, specifically the binding energy, the charge radius, and the AC (see Ref. 52)

$$V_0 = 363.351572 \text{ MeV and } \alpha = 0.4 \text{ fm}^{-2}. \qquad (9)$$

The binding energy of –6.812200 MeV, the mean square charge radius of 2.53 fm, and the mass radius of 2.54 fm were obtained with this potential. The experimental value for the charge radius of $^{10}$Be is absent in Refs. 52-54 and for $^9$Be it equals 2.518(12) fm.[54] Later we will consider that the $^{10}$Be radius should not have a large excess over the $^9$Be radius. We are assuming that the charge neutron radius equals zero, because small deviations from zero do not play a principal role here. Its mass radius is equal to the proton radius of 0.8775(51) fm.[31] The asymptotic constant, calculated according to the Whittaker functions,[40] is equal to $C_W = 1.73(1)$ at the distance 4–16 fm. The AC error is obtained by averaging it over the stated interval, where the asymptotic constant remains practically stable. Besides the allowed BS corresponding to the ground state of $^{10}$Be with {4411}, such $^3P_0$ potential has the FS with {541} in full accordance with the second variant of the classification of orbital states of clusters in the system of 10 particles in the 9+1 channel given above.

Let us cite the results of Ref. 55 for comparison of the asymptotic constants, where $C^2 = 1.69(15)$ fm$^{-1}$ was obtained. It is necessary to note that for the definition of the asymptotic constant in that work the expression $\chi_L(r) = CW_{-\eta L+1/2}(2k_0 r)$ was used, but in our calculations the expression of Eq. (7).[30,40] The result of the AC from Ref. 55 is obtained for a neutron spectroscopic $S_n$ factor that is not equal to unity, as it is accepted in the present calculations. Therefore, the initial value should be divided by the value $S_n$ in this channel,[55] which is equal to 0.2 according to results of theoretical work.[56] However, the value of 0.92 is given in the review of Ref. 52 from the analysis of experimental data of the $^9$Be($^2$H, p$_0$)$^{10}$Be reaction for $S_n$ of $^{10}$Be in the n$^9$Be channel. Since we did not find other results for the value of $S_n$, it will be logical to accept that its average value is equal to 0.56, which in dimensionless form with $\sqrt{2k_0} = 1.045$ for the AC[55] in our definition takes $C_W = 1.66(7)$. Thus, it can be considered that the GS potential of $^{10}$Be in the n$^9$Be channel was constructed on the basis of description of the binding energy in this cluster channel and the channel AC value.

The VM with the expansion of the cluster wave function of the relative motion of the n$^9$Be system on a non-orthogonal Gaussian basis[10,30] was used for additional control of the binding energy calculations of Eq. (6). With the dimension of the basis $N = 10$, an energy of –6.812193 MeV was obtained for the ground state potential of Eq. (9); this result differs from the FDM value given above by only 7.0 eV.[10] The residuals are of the order of 10$^{-10}$, the asymptotic constant in the range 5–12 fm is equal to 1.73(2), and the charge radius does not differ from the previous results.[10]

Since the variational energy decreases as the dimension of the basis increases and yields the upper limit of the true binding energy, and the finite-difference energy increases as the step size decreases and number of steps increases,[54] the average value of –6.8121965(35) MeV can be taken as a realistic estimate of the binding energy in this potential. Therefore, the real accuracy of determination of the binding energy of $^{10}$Be in the n$^9$Be cluster channel for this potential, using two different methods (FDM and VM) and two different computer programs for the potential of Eq. (9), is at the level of ±3.5 eV.

For the potential of the first $^3P_2^1$ excited state (ES) of $^{10}$Be at the energy of –3.44417 MeV in the n$^9$Be channel with $J^\pi = 2^+$ the parameters are obtained as:

$$V_0 = 345.676472 \text{ MeV and } \alpha = 0.4 \text{ fm}^{-2}. \qquad (10)$$

This potential results in an energy of –3.444170 MeV, a charge radius of 2.54 fm, a mass radius of 2.57 fm, and an AC equal to 1.15(1) in the interval of 4–18 fm, and has one bound FS at {541}; i.e., it applies to the given classification of the orbital cluster states according to Young tableaux. The state with {4411} is considered here as unbound. Since we have no data on ACs of the ESs, then the width of these potentials $\alpha$ is used for these potentials, which for the GS potential provides the true value of its AC.

The next parameters,

$$V_0 = 328.584413 \text{ MeV and } \alpha = 0.4 \text{ fm}^{-2}, \qquad (11)$$

were obtained for the potential of the second ES $^3P_2^2$ of $^{10}$Be at the energy of –0.85381 MeV in the n$^9$Be channel with $J^\pi = 2^+$, which also has one FS. This potential allows us to obtain a binding energy of –0.853810 MeV, a charge radius of 2.55 fm, a mass radius of 2.69 fm, and an AC equal to 0.60(1) in the interval of 4–26 fm.

Furthermore, let us note that the fourth $^3P_0^2$ excited level at the energy of 6.1793 MeV relative to the GS or –0.6329 MeV relative to the threshold of the n$^9$Be channel of $^{10}$Be coincides with the GS with respect to its quantum numbers $J^\pi T = 0^+ 1$.[52] Therefore, it is possible to consider the $E1$ transition to this BS from the $S$ scattering wave. The potential of this BS has the parameters:

$$V_0 = 326.802239 \text{ MeV and } \alpha = 0.4 \text{ fm}^{-2}, \qquad (12)$$

and leads to a binding energy in this channel of –0.632900 MeV, a charge radius of 2.56 fm, a mass radius of 2.72 fm, and an AC equal to 0.53(1) in the interval of 4–28 fm.

Besides, the third $^3S_1$ excited level with $J^\pi T = 1^- 1$ (see Ref. 52) is observed at the energy of 5.9599 MeV or –0.8523 MeV relative to the threshold of the n$^9$Be channel. Therefore, it is possible to consider the $M1$ transition from the $^3S_1$ non-resonance scattering wave to the same $^3S_1$ bound state of $^{10}$Be. However, it is usually considered that the cross-sections of such a $^9$Be(n,$\gamma_3$)$^{10}$Be process will be lower by one to two orders of magnitude than those obtained in the $E1$ capture.

Pay attention again that, similarly to the n$^{12}$C system,[44] the third ES of $^{10}$Be with $J^\pi = 1^-$ at 5.9599 MeV could be considered as the bound AS in the $^3S_1$ scattering wave with Young tableau {422}. Therefore, we use the following potential for the $^3S_1$ wave:

$$V_0 = 33.768511 \text{ MeV and } \alpha = 0.4 \text{ fm}^{-2}, \qquad (13)$$

which leads to a binding energy of –0.852300 MeV,[52] charge and mass radii of 2.57 fm and 2.76 fm, and an AC equal to 1.24(1) in the interval of 4–30 fm. The phase shift of this potential is shown in Fig. 1 by the dashed line and smoothly decreases till 121° at 1.0 MeV.

Such potential does not contain the FS and will be used for calculations of the total cross sections of the $E1$ transition of the form ${}^3P_0 + {}^3P_1 + {}^3P_2 \to {}^3S_1$ and from the ${}^3S_1$ scattering wave with the potential of Eq. (13) to the GS ${}^3P_0^1$ of ${}^{10}$Be in the n${}^9$Be channel and the fourth ES ${}^3P_0^2$ at 6.1793 MeV with $J^\pi = 0^+$, i.e., ${}^3S_1 \to {}^3P_0^1 + {}^3P_0^2$. In addition, the transition to the first ES at the energy of 3.36803 MeV and to the second ES ${}^3P_2^2$ at the energy of 5.95839 MeV with $J^\pi = 2^+$, i.e., ${}^3S_1 \to {}^3P_2^1 + {}^3P_2^2$ will be considered.

Let us consider fifth excited level with $J^\pi T = 2^-1$ and energy of 6.26330 MeV, bounded at –0.5489 MeV relatively to the threshold of the n${}^9$Be channel of ${}^{10}$Be.[52] It can be compared to ${}^3D_2$ wave for triplet spin state; therefore the $E1$ transition of the form ${}^3P_0 + {}^3P_1 + {}^3P_2 \to {}^3D_2$ is possible. Consequently, the next parameters were obtained for the ${}^3D_2$ potential:

$$V_0 = 530.56378 \text{ MeV and } \alpha = 0.4 \text{ fm}^{-2}, \quad (14)$$

This potential contains the FS and leads to a binding energy of –0.548900 MeV, charge radius of 2.54 fm, an AC equal to 0.080(1) in the interval 4–13 fm, and its phase shift at 1.0 MeV decrease down to 359°. Thus, this potential has forbidden state, then the phase shifts obey to the Levinson theorem[5] and beginning from 360°.

Furthermore, let us consider the resonance states of the n${}^9$Be scattering at the energy lower than 1 MeV. It is known that in the spectrum of ${}^{10}$Be in the n${}^9$Be channel there is superthreshold level with $J^\pi = 3^-$ at the energy of 0.6220 MeV in the laboratory system (l.s.) and width of 15.7 keV in the center-of-mass system (c.m.),[52] which in the triplet state can be matched to the ${}^3D_3$ resonance in the n${}^9$Be elastic scattering. The level with $J^\pi = 2^+$ at the energy of 0.8118 MeV (l.s.) and width of 6.3 keV (c.m.), which can correspond to the ${}^3P_2$ resonance in the n${}^9$Be scattering[52,53] is also observed. In the first case, the $E3$ transition has to be considered and in the second the $M2$ transition to the GS of ${}^{10}$Be, whose cross-sections are appreciably lower than those of the $E1$ process and will not be taken into account in future.

However, it is possible to consider the $E1$ transition, for example, from the ${}^3D_3$ resonance scattering wave to the first and second excited states of ${}^{10}$Be with $J^\pi = 2^+$ at the energies of 3.36803 MeV and 5.95839 MeV relative to the GS, which are bound at –3.44417 MeV and –0.85381 MeV relative to the threshold of the n${}^9$Be channel,[52] i.e., ${}^3D_3 \to {}^3P_2^1 + {}^3P_2^2$.

The next parameters were found for the ${}^3D_3$ scattering wave

$$V_0 = 457.877 \text{ MeV and } \alpha = 0.35 \text{ fm}^{-2}. \quad (15)$$

The phase shift of this potential is shown in Fig. 1 by the solid line and has a resonance character, and the potential itself contains the bound FS with {541} tableau in accordance with the classification given above, and the state with {442} is not bound. This variant of the potential, with this structure of FSs, will be considered as the main, but for comparison we will further consider two other options.

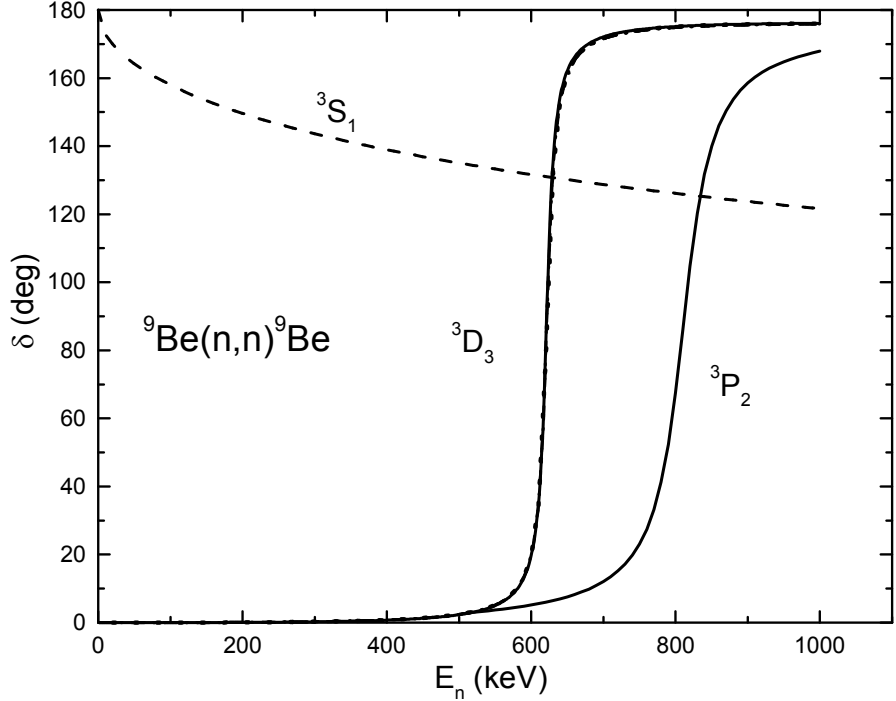

Fig. 1. The $^3S_1$, $^3P_2$ and $^3D_3$ phase shifts of the n$^9$Be elastic scattering at low energies. The lines show the calculations with the Gaussian potential, whose parameters are given in the text. The *D* and *P* phase shifts with FSs are shifted down from 180° to 0°.

If one uses the expression[57] for the calculation of the level width by the phase shift $\delta$,

$$\Gamma = 2(d\delta/dE)^{-1}, \quad (16)$$

then the width of this resonance is equal to 15.0 keV (c.m.), which is in quite good agreement with the results of work.[52] The second variant of the potential has the parameters

$$V_0 = 132.903 \text{ MeV and } \alpha = 0.22 \text{ fm}^{-2}. \quad (17)$$

The phase shift of this potential is shown by the dot-dashed line in Fig. 1, the resonance is also at 622 keV with a width of 15.7 keV, and the potential itself does not contain bound forbidden or allowed states; i.e., this potential corresponds to the case when FS for {541} is absent, and the state with {442} is not bound.

Now, we will consider the option of the $^3D_3$ scattering potential with the parameters:

$$V_0 = 985.183 \text{ MeV and } \alpha = 0.43 \text{ fm}^{-2}, \quad (18)$$

which leads to the width of 15.5 keV at the resonance of 622 keV and has two bound states. First of them corresponds to the bound FS with {541}, the second to the bound state at {442}. The scattering phase shift is shown in Fig. 1 by the dotted line, which differs from the solid line only at energies lower than 400 keV.

The next parameter values are used for the potentials of the $^3D_2$ and $^3D_1$ waves:

$$V_0 = 300.0 \text{ MeV and } \alpha = 0.35 \text{ fm}^{-2}, \qquad (19)$$

which lead to the zero phase shifts, because there are no resonances in these waves. This potential contains the bound FS with {541} in accordance with the given above classification, and the state {442} is not bound.

As said above, there is the resonance level in the n$^9$Be elastic scattering with $J^\pi = 2^+$ at the energy of 0.8118(7) MeV (l.s.) above the threshold of the n$^9$Be channel and with the width of 6.3 keV (c.m.) that can be matched to the $^3P_2$ resonance.[52] We have not had much success in finding the potential for this resonance, which is able to correctly describe its small width of 6.3 keV,[52] and for this wave the next parameters have been obtained:

$$V_0 = 32072.526 \text{ MeV and } \alpha = 40.0 \text{ fm}^{-2}. \qquad (20)$$

The potential leads to the resonance at 812 keV with the phase value of 90.0°(5), width of 53 keV, contains one FS and its phase shift is shown in Fig. 1 by the solid line in the range of 0.8 MeV. Once more notice that using the given number of BSs, FSs and ASs, the position of the resonance and its width, the partial potential is constructed completely unambiguously. Therefore, we would have created more narrow potential for obtaining correct width of the resonance, but it already has unusual parameters of the depth and the width, because of presence of the FS. We will use this potential furthermore for considering the $E1$ transition from the resonance $^3P_2$ scattering wave to the third $^3S_1$ excited state, but bound in the n$^9$Be channel of $^{10}$Be. Generally, the next transitions are possible here: $^3P_0 + {}^3P_1 + {}^3P_2 \to {}^3S_1$.

The potential of the partial $^3P_0$ scattering wave is taken in the form of Eq. (9). It leads to the smoothly decreasing phase shifts from 360° down to 355° at 1.0 MeV. There are no resonances and excited bound states in the n$^9$Be channel in the partial $^3P_1$ scattering wave. Therefore, such potentials containing one bound FS for tableau {541}, but without bound AS for {4411}, have to give near zero scattering phase shift. The next parameters are obtained for it:

$$V_0 = 206.0 \text{ MeV and } \alpha = 0.4 \text{ fm}^{-2}. \qquad (21)$$

This potential gives the scattering phase shift in the range of 180.0°(1) at the energy up to 1.0 MeV. Thereby, the same potentials of the n$^9$Be interaction are used in the $^3S_1$, $^3D_2$, and $^3P_0$ waves in continuous and discrete spectra, and the potentials $^3P_0$, $^3P_1$ and $^3P_2$ take into account the absence (at $J^\pi = 0^+$ and $1^+$), or presence (at $J^\pi = 2^+$) of the resonances in these partial waves.

Accurate values of particle masses are specified in our calculations;[31,58] the $\hbar^2/m_0$ constant, where $m_0$ is the atomic unit, is equal, as usual, to 41.4686 MeV fm$^2$. Although it is currently considered that this value is slightly out of date, we are continuing to use it for ease of comparison of the new results with all previous results (see, for example, Refs. 5, 7, 30 and 59]).

### 3.3. *Total capture cross-sections*

The expressions for the total radiative capture cross-sections $\sigma(NJ,J_f)$ in the potential cluster model are given above in Eq. (3). The $E1$ process from the $^3S_1$ non-resonance scattering wave with the potential of Eq. (13) to the $^3P_0^1$ bound ground state of $^{10}$Be with $J^\pi T = 0^+1$ for the potential of Eq. (9), i.e., $^3S_1 \to\, ^3P_0^1$ transition, was taken into account in the consideration of the electromagnetic transitions in the n$^9$Be cluster channel. The calculation results of the total capture cross-sections are shown in Fig. 2a by the dot-dot-dashed line for these potentials. The experimental data for the total cross-sections at energies from 25 meV to 25 keV (l.s.) were taken from Refs. 60, 61 and 62, respectively.

The dashed line in Fig. 2a gives the results for the total cross-sections of the transition $^3S_1 \to\, ^3P_2^1$ with the combination of the potentials of the first ES in the $^3P_2$ wave of Eq. (10) and the $^3S_1$ scattering wave of Eq. (13). The dot-dashed line shows the results for the transition $^3S_1 \to\, ^3P_2^2$ with potentials of the second ES of Eq. (11) and the $^3S_1$ wave of Eq. (13). The dotted line shows the results for the transition from the $^3S$ scattering wave of Eq. (13) to the fourth ES $^3P_0^2$ with $J = 0^+$ with the potential of Eq. (12), and the solid line is the sum of all these transitions.

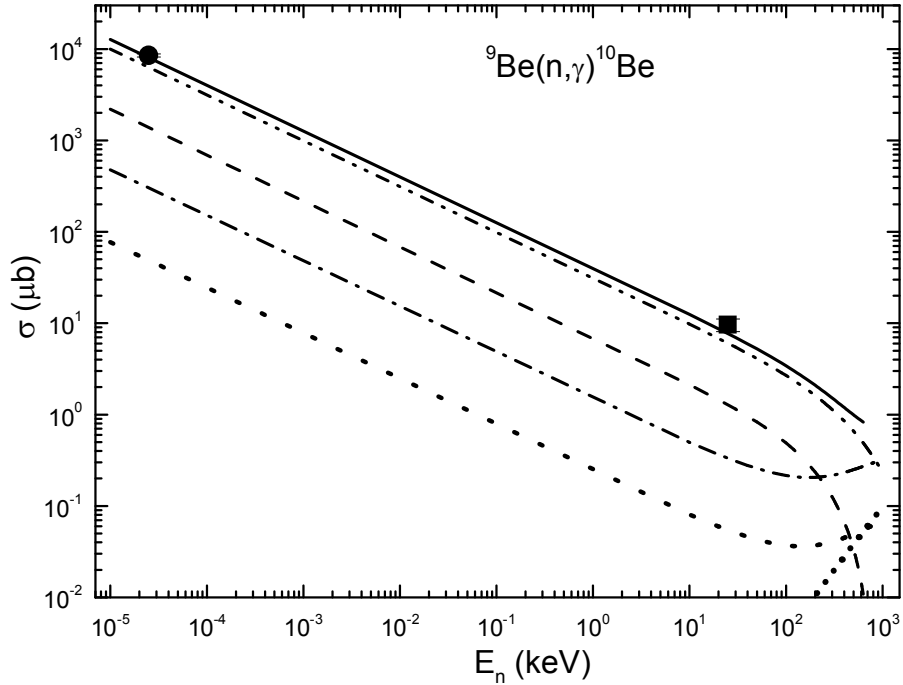

Fig. 2a. The total cross-sections of the radiative neutron capture on $^9$Be. Points (●) indicate experimental data from Ref. 60 at 25 meV and squares (■) indicate experimental data from Refs. 61 and 62 at 25 keV. Lines show the total cross-sections calculation results.

Also, we have considered the transitions from the $^3D_2$ or $^3D_1$ scattering states with potentials of Eq. (18) to the first ES with maximal binding energy for the potential of Eq. (10), i.e., $^3D_1 +\, ^3D_2 \to\, ^3P_2^1$. Moreover the transition from the $^3D_1$ scattering state to the GS of $^{10}$Be with the potential of Eq. (9), i.e., $^3D_1 \to\, ^3P_0^1$ was considered. In the first case, the capture cross section is shown by short dots in the right part of Fig. 2a. There is practically no difference from the results of second case with the transition to the GS. Hence it follows that such transitions become play a visible role only at energies up to 1 MeV.

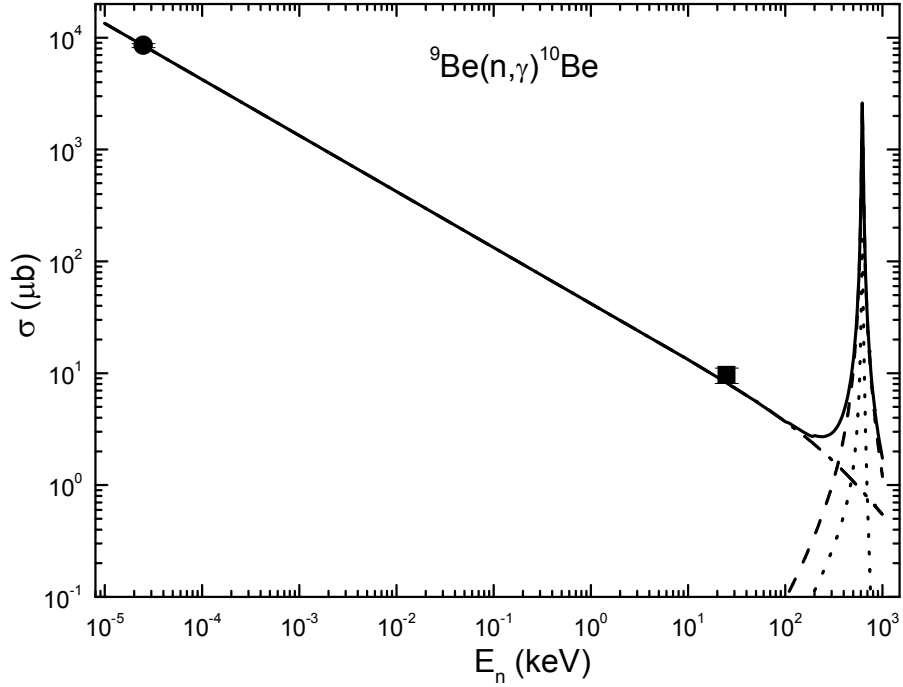

Fig. 2b. The total cross-sections of the radiative neutron capture on $^9$Be taking into account the resonance at 622 keV. Points (●) represent experimental data from Ref.60 at 25 meV and squares (■) represent experimental data from Refs. 61 and 62 at 25 keV. Lines show the total cross-sections calculation results.

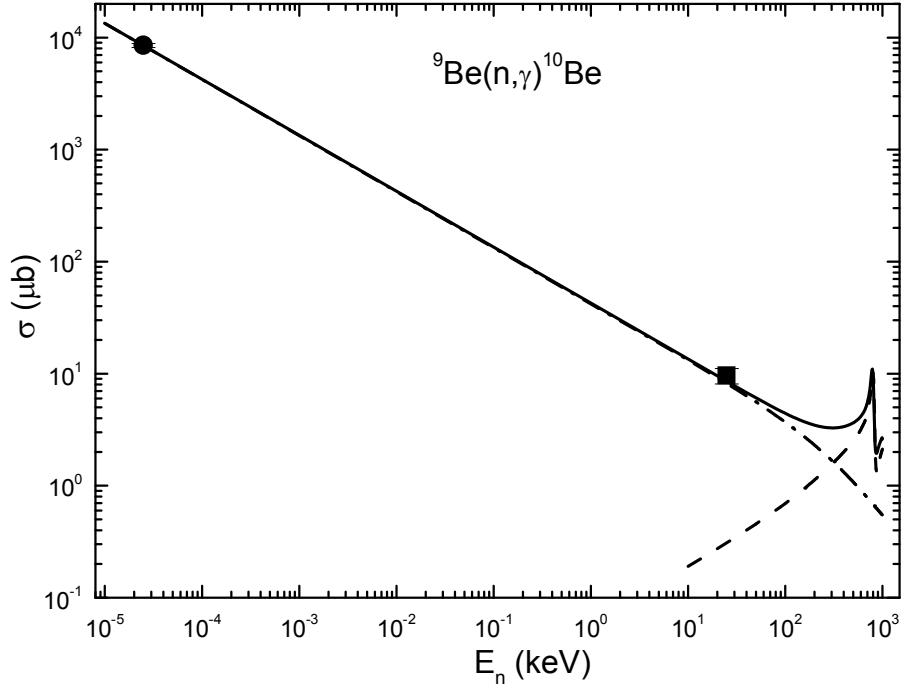

Fig. 2c. The total cross-sections of the radiative neutron capture on $^9$Be taking into account the resonance at 812 keV. Points (●) represent experimental data from Ref. 60 at 25 meV and squares (■) represent experimental data from Refs. 61 and 62 at 25 keV. Lines show the total cross-sections calculation results.

Furthermore, we have considered the $E1$ transitions from the $^3D_3$ resonance scattering wave with $J^\pi T = 3^-1$ at the energy of 0.622 MeV (l.s.) and the width of 15.7 keV[52] to the first and second excited states of $^{10}$Be with $J^{\pi}T = 2^+1$, which can be compared with $^3P_2$ exited states of the n$^9$Be system. The calculation results of the total cross-section of the E1 transition from $^3D_3$ scattering wave to the second ES ($D_3 \to P_2^2$) with potentials of Eqs. (11) and (15) are shown in Fig. 2b by the dotted line, while the dashed line denotes the transition to the first ES ($D_3 \to P_2^1$) with potentials of Eqs. (10) and (15).

The dot-dashed line gives the sum of all cross-sections of the E1 transition, shown in Fig. 2a by the solid line. The solid line in Fig. 2b is the sum of all cross-sections of the considered transitions in the $^9$Be(n, γ)$^{10}$Be reaction. The total cross-section at the resonance energy of 622 keV from the $^3D_3$ scattering wave to the first excited state at the energy of –3.44417 MeV reaches 2.45 mb and that to the second one at the energy of –0.85381 MeV approximately equals 0.16 mb, at the summed value of ~2.6 mb. This value can be used for the estimation of the cross-section in the resonance at 622 keV, if such experimental measurements will be made.

The results of calculations of $^9$Be(n, γ$_3$)$^{10}$Be for the sum of all three E1 transitions $^3P_0 + ^3P_1 + ^3P_2 \to ^3S_1$ with potentials of all P waves of Eqs. (9), (20), (21) and ES of Eq. (13) are given in the right part of Fig. 2c by the dashed line. The results of calculations of $^9$Be(n, γ$_5$)$^{10}$Be for the sum of all three E1 transitions $^3P_0 + ^3P_1 + ^3P_2 \to ^3D_2$ with potentials of all P waves of Eqs. (9), (20), (21) and ES of Eq. (14) give the cross sections about 0.5 mb and practically not influence to the summed resonance cross sections.

The dot-dashed line gives the sum of all cross-sections of the E1 transition, shown in Fig. 2a by the solid line. The solid line in Fig. 2c is the sum of all cross-sections of the considered here transitions in the $^9$Be(n, γ)$^{10}$Be reaction. Let us note that calculated cross sections at the resonance 812 keV have the value lower and the width more than real, because the using potential of Eq. (20) leads to the overestimated width of this resonance. Comparing Fig. 2b with Fig. 2c one can see that taking into account the transitions $^3P_0 + ^3P_1 + ^3P_2 \to ^3S_1$ with potential of Eq. (20) leads to the cross sections at the resonance 812 keV about 10 μb, and practically not influence to the total summed resonance cross sections of the transitions at the range of 0.2–1.0 MeV.

Since at the energies from 10$^{-5}$ keV up to 10 keV the calculated cross-section shown in Fig. 2a by the solid line is almost a straight line, it can be approximated by a simple function of the form:

$$\sigma_{ap} = \frac{A}{\sqrt{E_n}}. \qquad (22)$$

The value of constant $A = 40.2008$ μb·keV$^{1/2}$ was determined from a single point of the cross-sections with a minimal energy of 10$^{-5}$ keV. Furthermore, it is possible to consider the absolute value of the relative deviation of the calculated theoretical cross-sections and the approximation of this cross-section by the expression given above in the energy range from 10$^{-5}$ to 10 keV:

$$M(E) = \left|[\sigma_{ap}(E) - \sigma_{theor}(E)]/\sigma_{theor}(E)\right|. \qquad (23)$$

It was found that this deviation is at the level of 0.7% at energies lower than 10 keV, making it possible to use the cross-section approximation given above in the majority of applied problems. It appears to be possible to suppose that this form of total cross-section dependence on energy will have conserved at lower energies. In this case, estimation of the cross-section value, for example at the energy of 1 µ keV, gives the value of 1.3 b.

## 4. Radiative neutron capture on $^{14}$C and $^{14}$N in cluster model

Starting the analysis of total cross sections of the neutron capture on $^{14}$C and $^{14}$N with the formation of $^{15}$C and $^{15}$N in the GS, let us note that the classification of orbital states of $^{14}$C in the n$^{13}$C system or $^{14}$N in the p$^{13}$C channel according to Young tableaux was considered by us earlier in works of Refs. 44 and 63. However, we regard the results on the classification of $^{15}$C and $^{15}$N nuclei by orbital symmetry in the n$^{14}$C and n$^{14}$N channels as the qualitative ones as there are no complete tables of Young tableaux productions for the systems with a number of nucleons more than eight,[47] which have been used in earlier similar calculations.[6,7,64] At the same time, just on the basis of such classification, we succeeded with description of available experimental data on the radiative capture processes: $^{13}$C(n, γ)$^{14}$C (Refs. 44, 64) and $^{13}$C(p, γ)$^{14}$N (Ref. 63). Therefore, here we will use the classification procedure of cluster states by orbital symmetries, which leads us to certain number of FS and AS in partial intercluster potentials, so, to certain number of nodes for wave function of cluster relative motion.

Furthermore, let us suppose that it is possible to use the orbital Young tableau {4442} for $^{14}$C ($^{14}$N), therefore we have {1} × {4442} → {5442} + {4443} for the n$^{14}$C system within the frame of 1p shell.[47] The first of the obtained tableaux compatible with orbital moments $L = 0, 2$ and is forbidden, because there can not be five nucleons in the s shell, and the second tableau is allowed and compatible with the orbital moment $L = 1$.[46] Thus, limiting oneself only by lowest partial waves with orbital moments $L = 0$ and 1, it is possible to say that there are forbidden and allowed states in the $^2S_{1/2}$ potential for the n$^{14}$C system. The last of them corresponds to the GS of $^{15}$C with $J^\pi = 1/2^+$ and lays at the binding energy of the n$^{14}$C system –1.21809 MeV.[65] At the same time the potential of the $^2P$ elastic scattering waves has not FS. In the case of the n$^{14}$N system, the forbidden state is in the $^2S_{1/2}$ elastic scattering wave, and the $^2P_{1/2}$ wave has only AS, which is at the binding energy of the n$^{14}$N system –10.8333 MeV.[65]

### 4.1. *Total cross sections of the neutron capture on $^{14}$C*

Now we will consider the radiative neutron capture process on $^{14}$C at energies from 20 keV to 1 MeV, approximately, where there are experimental data that are taken from the Moscow State University (MSU) data base.[58] The value obtained on the basis of resonance data at 3.103(4) MeV relative to the GS of $^{15}$C with $J^\pi=1/2^-$, i.e., approximately higher by 1.9 MeV (c.m.) than threshold of the n$^{14}$C channel and with the width about 40 keV[65] can be used for the $^2P_{1/2}$ wave potential of the n$^{14}$C scattering without FS. Because, we are studying only the energy range not far than 1.0 MeV, so it can be considered that the $^2P_{1/2}$ scattering phase shift simply equals zero. Consequently, the potential depth $V_0$ can be equaled to zero, because it has not forbidden states. The same applies to the $^2P_{3/2}$ scattering wave potential too, since the relevant resonance of $^{15}$C nucleus lays at higher energy value of 4.66 MeV.

The potential of the $^2S_{1/2}$ bound state with one FS has to correctly reproduce ground state binding energy of $^{15}$C with $J^\pi = 1/2^+$ in the n$^{14}$C channel at -1.21809 MeV[65] and reasonably describes the mean square radius of $^{15}$C, which value, apparently, should not considerably exceed the radius of $^{14}$C that is equal to 2.4962(19) fm.[65] Consequently, the next parameters were obtained:

$$V_{GS} = 93.581266 \text{ MeV}, \quad \alpha_{GS} = 0.2 \text{ fm}^{-2}. \qquad (24)$$

The potential leads to the binding energy of $-1.2180900$ MeV at the accuracy of using finite-difference method (FDM)[66] equals $10^{-7}$ MeV, to the mean square charge radius $R_{ch} = 2.52$ fm and mass radius of 2.73 fm. The zero value was used as neutron charge radius and its mass radius has taken equal to proton radius.[31] The value of 1.85(1) was obtained for the asymptotic constant at the range 7–27 fm. The error of the constant is determined by its averaging over the above mentioned distance.

The value of 1.13 fm$^{-1/2}$ for the AC was given in Ref. 27 with the reference to Ref. 67, that gives 1.65 after recalculations to the dimensionless quantity at $\sqrt{2k} = 0.686$. The detailed review of the values of this constant is given in one of the latest work of Ref. 68 devoted to the determination of the AC from the characteristics of different reactions – its value is in the limit from 1.22(6) fm$^{-1/2}$ to 1.37(5) fm$^{-1/2}$, that gives 1.8–2.0 after recalculations. In Ref. 68 itself, the values of AC were obtained from 1.25 fm$^{-1/2}$ to 1.52 fm$^{-1/2}$ or after recalculations in the interval 1.8–2.2 with the recommended value 1.87(6), which practically does not differ from the value that we have obtained.

The variational method[66] was used for the additional control of calculations of the GS energy of $^{15}$C, which with the dimension of the basis $N = 10$ and independent variation of the potential parameters for the BS potential of Eq. (24) allowed to obtain the energy of $-1.2180898$ MeV. The asymptotic constant at the range of 10–25 fm and the charge radius do not differ from the previous results of the FDM. Since, the variational energy decreases as the dimension of the basis increases and yields the upper limit of the true binding energy,[64] and the finite-difference energy increases as the step size decreases and number of steps increases,[66] the average value of $-1.2180899(1)$ MeV can be taken as a realistic estimate of the binding energy in this potential. Therefore, the real accuracy of determination of the GS binding energy of $^{15}$C in the n$^{14}$C cluster channel for the potential of Eq. (24), using two methods (FDM and VM) and two different computer programs, is at the level of $\pm 0.1$ eV.

Going to the results of our calculations, let us note that available experimental data for the total cross sections of the radiative neutron capture on $^{14}$C,[69,70,71,72,73] found using the MSU data base,[58] show the existence of big ambiguities of these cross sections, measured in different works. For example, the difference of cross sections at the energy 23 keV[70–72] is equal to two-three times, and ambiguity of the different data at the energy range 100–1000 keV reaches three-four times.[69,71–73] The experimental results at the energy range 23 keV – 1.0 MeV for works mentioned above are shown in Figs. 3a and 3b.

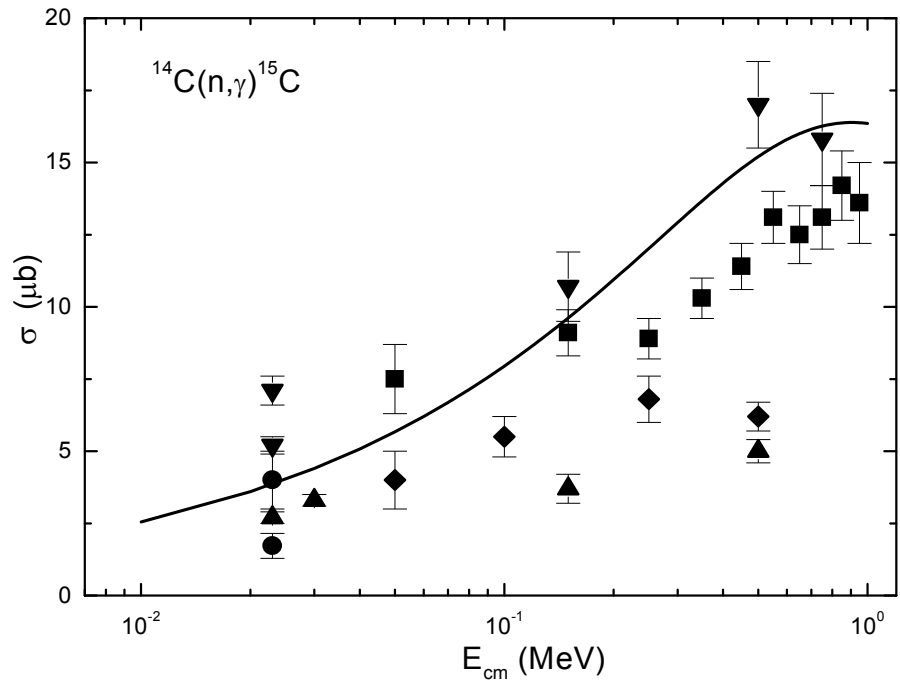

Fig. 3a. The total cross sections of the radiative neutron capture on $^{14}$C. Squares (■) represent experiment data from Ref. 69, points (●) – from Ref. 70, triangles (▲) – from Ref. 71, reverse triangles (▼) – from Ref. 72, rhombs (♦) – from Ref. 73. Lines show results of calculations of the total cross-sections.

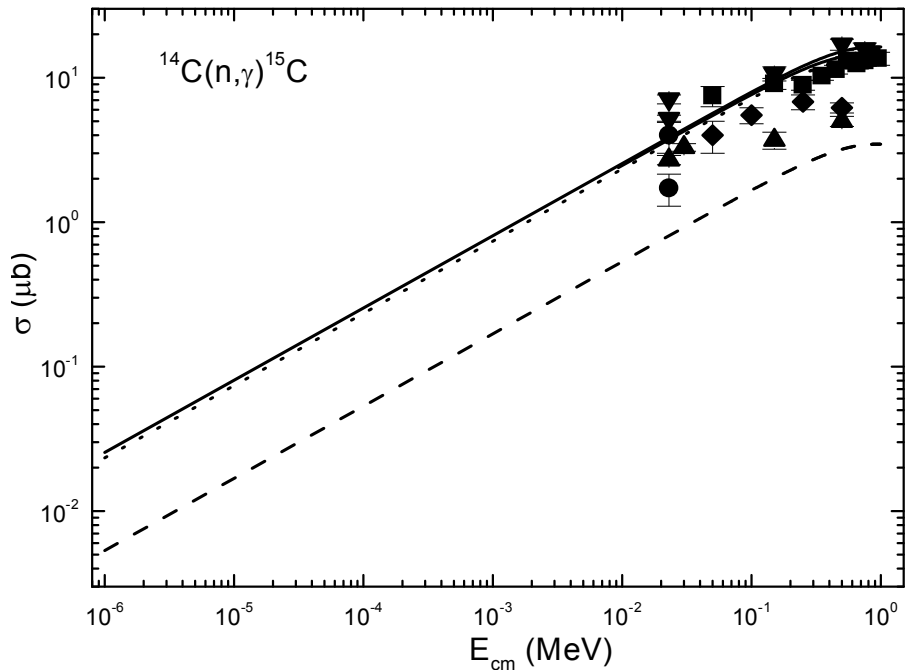

Fig.3b. The total cross sections of the radiative neutron capture on $^{14}$C. Squares (■) represent experiment data from Ref. 69, points (●) – from Ref. 70, triangles (▲) – from Ref. 71, reverse triangles (▼) – from Ref. 72, rhombs (♦) – from Ref. 73. Lines show results of calculations of the total cross sections.

Here, we are considering only the $E1$ capture from the $^2P$ scattering states to the $^2S_{1/2}$ ground state of $^{15}$C, because the contribution of the $E2$ radiative capture process with the transition from the $^2P$ scattering waves to the first excited state with $J^\pi = 5/2^+$ that can be compare to the $^2D_{5/2}$ level is 25–30 times less as it was shown in Ref. 74, and it is possible to neglect it in the presence of existent errors and ambiguities in measurements of total cross sections.

Thereby, we have considered only the $E1$ transition to the GS from the non-resonance $^2P_{1/2}$ and $^2P_{3/2}$ scattering waves at energies lower 1 MeV, with zero depth potential without FS, i.e., with zero phase shifts. The total cross section results of calculations of the radiative neutron capture on $^{14}$C with the GS potential of Eq. (24) given above, at energies lower 1 MeV, are shown in Figs. 3a and 3b by the solid lines, at that the results of calculations from Fig. 3b are starting from the energy 1 eV. It is seen that these results are in a better agreement with data from Ref. 72. Thereby, the total capture cross sections completely depend from the form of the potential of the ground state of $^{15}$C in the n$^{14}$C channel, because the $^2P$ potentials of the input channel without FSs can be simply zeroized at the considered energies. The used GS potential, which allows us to acceptably describe the existent experimental data for the total radiative capture cross sections, leads to the correct description of the basic characteristics of the GS, specifically, the binding energy, the mean square radius and the AC of $^{15}$C in the n$^{14}$C channel.

Let us note that if we will use the GS potential of $^{15}$C without FS, for example, with parameters:

$$V_{GS} = 19.994029 \text{ MeV and } \alpha_{GS} = 0.2 \text{ fm}^{-2}, \qquad (25)$$

leading to the binding energy of -1.218090 MeV, the AC of 1.46(1) at the range of 5–30 fm, the charge and mass radiuses of 2.51 fm and 2.63 fm, correspondingly. The total cross section calculation results are shown in Fig. 3b by the dashed line, which is appreciably lower than all experimental data.

For the purpose of obtaining results that, in general, correctly describe experiment, we need parameters of the potential without FS:

$$V_{GS} = 4.593639 \text{ MeV and } \alpha_{GS} = 0.02 \text{ fm}^{-2}, \qquad (26)$$

that leads to the big width and small depth of the interaction; the cross section calculations are shown in Fig. 3b by the dotted line, which is practically superposed with the solid line. This potential leads to the charge radius of 2.53 fm, the mass radius of 2.92 fm, and its AC at the interval of 15–30 fm is equal to 3.24(1). We can see from these results that both variants of these GS potentials reproduce the AC value[68] incorrectly, although the last of them allows to reproduce the experimental total capture cross sections.

Because at energies from 1 eV up to 1 keV the calculated cross section is almost a straight line (see Fig. 3b, solid line), it can be approximated by a simple function of the form:

$$\sigma_{ap} = 0.7822\sqrt{E_n} . \qquad (27)$$

The value of given constant 0.7822 μb·keV$^{-1/2}$ was determined by a single point at a cross section with minimal energy of 1·eV. Furthermore, it turns out that the absolute value of the relative deviation of the calculated theoretical cross sections, and the approximation of this cross section by the expression given above at energies less than 1 keV, does not exceed 0.4%. Apparently, it can be suggested that this form of the total cross section energy dependence will be retained at lower energies. Therefore, we can perform an evaluation of the cross section value; for example, at the energy of 1 meV – it gives result 0.78 $10^{-3}$ μb.

Thus, it is possible to describe the available experimental data only on the basis of the $E1$ transition from the $P$ scattering waves with zero potentials to the GS of $^{15}$C, which is described by the interactions that coherent with the main characteristics of this state – the mean square radius, the AC in the considered n$^{14}$C channel and the two-body binding energy. However, since for almost all earlier considered by us neutron capture processes on light nuclei, the total cross sections were measured down to 5–25 meV,[7,44] then it is interesting to carry out the measurements of the cross sections for this reaction, even though for energy about 1 eV, when the given above formula forecast the value of 0.025 μb.

### 4.2. *Total cross sections of the neutron capture on $^{14}$N*

The ground state of $^{15}$N with $J^\pi, T = 1/2^-, 1/2$, since $^{14}$N has the moment $J^\pi, T = 1^+, 0$,[65] can be represented by the mixture of doublet $^2P_{1/2}$ and quartet $^4P_{1/2}$ states and, further, the $E1$ transitions at the non-resonance energy range of 0.5–0.6 MeV from the doublet $^2S_{1/2}$ and quartet $^4S_{3/2}$ scattering waves with one bound FS to the $^{2+4}P_{1/2}$ GS – $^2S_{1/2} + ^4S_{3/2} \to ^{2+4}P_{1/2}^{gs}$ will be considered as basic, i.e., the next total capture cross sections will be calculated:

$$\sigma_0(E1) = \sigma(E1, ^2S_{1/2} \to ^2P_{1/2}^1) + \sigma(E1, ^4S_{3/2} \to ^4P_{1/2}^1). \tag{28}$$

Thereby, as for the neutron capture on $^7$Li, the transitions to the doublet and quartet parts of wave functions of the ground state, which are not differ in this approach and correspond to the BS in the same potential. It should be noted here that there is excited level in the $^{15}$N spectrum at the energy of 9.2221 MeV, but bound in the n$^{14}$N channel at -1.6112 MeV with $J^\pi = 1/2^-$ level. Thus, it is possible to analyze additional transitions $^2S_{1/2} + ^4S_{3/2} \to ^{2+4}P_{1/2}^{es}$, i.e., to consider total cross sections:

$$\sigma_1(E1) = \sigma(E1, ^2S_{1/2} \to ^2P_{1/2}^2) + \sigma(E1, ^4S_{3/2} \to ^4P_{1/2}^2). \tag{29}$$

Here, we will limit ourselves to just transitions on the bound states with minimal values of $J^\pi = 1/2^\pm$. Therefore, let us consider further the possible transitions from the $P_{1/2}$ and $P_{3/2}$ scattering states to the second, seven and nine ESs of $^{15}$N with $J^\pi = 1/2^+$ at the energies 5.298822, 8.31262 and 9.04971 MeV, binding in the n$^{14}$N channel, which can be refer to the $^2S_{1/2}$ doublet wave.

Meanwhile, furthermore we will consider the $^{2+4}P_{1/2}$ scattering state, which has the resonance at 492.6(0.65) keV and the width about 8(3) keV (see, for example, Tables 15.4 and 15.14 of Ref. 65 – the resonance at 11.2928(7) MeV). We will not take into account the resonance state placed near at 430(5) keV with $J^\pi \geq 3/2$ and width about 3 keV (energy of 11.235(5) MeV in Table 15.4 of Ref. 65), we will not taking into account, for a while, certain parity.[65] The potentials of the $^{2+4}P_{3/2}$ waves have to be equal to zero, because they have no FSs and resonances below 1.0 MeV. So, then we can analyze processes $^2P_{1/2} + ^2P_{3/2} \to ^2S_{1/2}^{es}$ and represent total cross sections in the form:

$$\sigma_3(E1) = \sigma(E1, ^2P_{1/2} \to ^2S_{1/2}^1) + \sigma(E1, ^2P_{3/2} \to ^2S_{1/2}^1) +$$
$$\sigma(E1, ^2P_{1/2} \to ^2S_{1/2}^2) + \sigma(E1, ^2P_{3/2} \to ^2S_{1/2}^2) + \quad (30)$$
$$\sigma(E1, ^2P_{1/2} \to ^2S_{1/2}^3) + \sigma(E1, ^2P_{3/2} \to ^2S_{1/2}^3).$$

Proceeding to the construction of the potentials for all these states, let us note that doublet and quartet bound $S$ levels with one bound FS are evidently differ in binding energy and the interaction potentials will be obtained for each of them. The $^2S$ and $^4S$ scattering potentials with one bound FS are also evidently differ, but we did not succeed in construction of the potentials, which are able to take into account the resonances in these waves. Therefore, further we will consider that they have to result in the near-zero scattering phase shifts and we will use the same potential parameters for them. Let us note, that the first resonance in the $^2S_{1/2}$ wave is at the energy of 11.4376(0.7) MeV with $J^\pi = 1/2^+$ or 0.639(5) MeV (c.m.) above threshold of the n$^{14}$N channel with the width of 34 keV (l.s.), and in the $^4S_{3/2}$ wave at the energy 11.763(3) MeV with $J^\pi = 3/2^+$ or 0.998(5) MeV above threshold of the n$^{14}$N channel with the neutron width about 45 keV (l.s.).[65]

The $^{2+4}P$ scattering states or the $P$ bound levels are mixed by spin, because the total moment $J^\pi = 1/2^-$ or $J^\pi = 3/2^-$ can be obtained both for $^2P$ and for $^4P$ waves. Therefore, further potentials of the $^{2+4}P_J$ states are constructed for states of the total moment $J$ and also become mixed by spin.

Totally, the parameter values that do not take into account the presence of the resonances in the considered energy range were used for the potentials of the doublet $^2S_{1/2}$ and quartet $^4S_{3/2}$ scattering waves

$$V_S = -19.0 \text{ MeV}, \quad \gamma_S = 0.06 \text{ fm}^{-2}, \quad (31)$$

and the results of calculations of the $^2S_{1/2}$ and $^4S_{3/2}$ phase shifts with this potential at the energies up to 1.0 MeV lead to the values in the range 0±2°.

The next parameters are used for the resonance potential at 493 keV of the $^{2+4}P_{1/2}$ wave without FS:

$$V_P = -13328.317 \text{ MeV}, \quad \gamma_P = 50.0 \text{ fm}^{-2}, \quad (32)$$

which lead to the resonance energy 493 keV at the level width 18.2 keV, that slightly more than the measured value.[65] Let us note that such potential should be done narrower for obtaining a correct value of the level width, while the width parameter already has the unusually big value.

The potential of the $^{2+4}P_{1/2}$ state without FS has to correctly reproduce the binding energy of the GS of $^{15}$N in the n$^{14}$N channel at -10.8333 MeV[65] and reasonably describe the mean square radius of $^{15}$N, which has the experimental value of 2.612(9) fm[65] at the experimental radius of $^{14}$N equals 2.560(11).[58] Consequently, the next parameters for the GS potential of $^{15}$N in the n$^{14}$N channel without FS were obtained:

$$V_{GS} = -55.442290 \text{ MeV}, \quad \gamma_{GS} = 0.1 \text{ fm}^{-2}. \tag{33}$$

The potential leads to the binding energy of -10.83330001 MeV at the FDM accuracy equals $10^{-8}$ MeV, to the mean square charge radius $R_{ch}$ = 2.57 fm and the mass radius of 2.73 fm. The value of 4.94(1) was obtained for the asymptotic constant in the dimensionless form at the range 7–13 fm. The error of the constant is determined by its averaging over the above mentioned distance. The value 5.69(7) fm$^{-1/2}$ is given in Refs. 75 and 76, and it gives 4.81(6) after recalculations to the dimensionless quantity at $\sqrt{2k} = 1.184$.

The variational method was used for the additional control of calculations of the GS energy, which with the dimension of the basis $N = 10$ and independent variation of the potential parameters for the potential of Eq. (33) allowed to obtain the energy of -10.83330000 MeV. The asymptotic constant at the range 7–14 fm is equal to 4.9(1), and the charge radius does not differ from the value obtained in above the FDM calculations.

Consequently, the average value of -10.833300005(5) MeV can be taken as a realistic estimate of the binding energy in this potential. Therefore, the accuracy of determination of the binding energy of $^{15}$N for the given two-cluster potential of Eq. (33) in the two-particle channel, obtained by two methods (FDM and VM) and by two different computer programs, can be written as $\pm 5 \cdot 10^{-9}$ MeV = $\pm 5$ meV and it practically coincide with the given FDM accuracy.

The potential parameters of the ES of $^{15}$N at the energy -1.6112 in the n$^{14}$N channel with the moment $J^\pi = 1/2^-$, coincided with the GS moment, have the values:

$$V_{ES} = -33.120490 \text{ MeV}, \quad \gamma_{ES} = 0.1 \text{ fm}^{-2}. \tag{34}$$

The potential leads to the binding energy of -1.611200 MeV at the FDM accuracy equals $10^{-6}$ MeV, to the mean square charge radius $R_{ch}$ = 2.58 fm and the mass radius of 2.71 fm. The value of 1.19(1) was obtained for the asymptotic constant in the dimensionless form at the range 8–30 fm.

The same width as for the GS was used for potentials of ESs with $J^\pi = 1/2^+$ of $^{15}$N, which compare to the $^2S_{1/2}$ bound levels in the n$^{14}$N channel. For example, for the potential of the first bound in the n$^{14}$N channel $^2S_{1/2}$ state at the energy 5.298822 MeV relatively to the GS or -5.534478 MeV relatively to the threshold of the n$^{14}$N channel the next values are used:

$$V_{1S} = -66.669768 \text{ MeV}, \quad \gamma_{1S} = 0.1 \text{ fm}^{-2}. \tag{35}$$

The potential leads to the binding energy of -5.534478 MeV, to the charge radius of 2.57 fm, the mass radius of 2.69 fm and the value of 5.65(1) was obtained for the AC at the range of 8–20 fm.

The next parameters were used for the potential of the $^2S_{1/2}$ second bound ES in the n$^{14}$N channel at the energy of 8.31263 MeV relatively to the GS or -2.52068 MeV relatively to the threshold of the n$^{14}$N channel:

$$V_{2S} = -56.271191 \text{ MeV}, \quad \gamma_{2S} = 0.1 \text{ fm}^{-2}. \tag{36}$$

The potential leads to the binding energy of -2.520680 MeV, to the charge radius of 2.58 fm, the mass radius of 2.77 fm and the value of 3.34(1) was obtained for the AC at the range of 8–21 fm.

These parameters were used for the potential of the third bound in the n$^{14}$N channel $^2S_{1/2}$ state at the energy of 9.04971 MeV relatively to the GS or -1.78362 MeV relatively to the threshold of the n$^{14}$N channel:

$$V_{3S} = -53.1402573 \text{ MeV}, \quad \gamma_{3S} = 0.1 \text{ fm}^{-2}. \tag{37}$$

The potential leads to the binding energy of -1.783620 MeV, to the charge radius of 2.58 fm, the mass radius of 2.82 fm and the value of 2.78(1) was obtained for the AC at the range of 8–27 fm.

Proceeding to the direct description of the results of our calculations, let us note that all used experimental data for total cross sections of the radiative neutron capture on $^{14}$N were obtained from the data base,[50,58] and the data itself are given in works.[77,78,79,80,81,82,83,84] These data for total cross sections of the radiative neutron capture on $^{14}$N have been obtained for the energy range from 25 meV to 65 keV and are shown in Fig. 4 together with our results of calculations (the solid line) of the total summarized cross sections of the radiative neutron capture on $^{14}$N to the considered above bound $^{2+4}P_{1/2}$ and $^2S_{1/2}$ states of $^{15}$N with the given potentials at energies lower than 1 MeV. The results of different works at energy 25 meV have the values of the cross sections at the range 77–80 mb. They are given in Fig. 4 by one point and, for example, in one of the latest work of Ref. 83 the value of 80.3(6) mb was obtained for this energy.

The calculations of transition from the S scattering waves of Eq. (31) to the GS of Eq. (33) are shown by the dashed line in the non-resonance range, the dashed-dot line shows the transition to the excited $^{2+4}P_{1/2}$ state with the potential of Eq. (34), and the dashed-dot-dot line identifies their sum. The resonance part of cross sections for potential of the $^2P_{1/2}$ scattering in the form of Eq. (32) and zero potential of the $^2P_{3/2}$ wave of continuous spectrum is formed by the transitions to the first, second and third bound $^2S_{1/2}$ states with potentials of Eqs. (35)-(37). These cross sections are denoted by the dashed, dotted and dashed-dot lines, and their sum is shown by the solid line in the resonance range, i.e., near 5–1000 keV. The total summarized cross sections taking into account all considered transitions are also shown by the solid line at all energy range, i.e., from $10^{-5}$ keV to 1.0 MeV.

It is seen from the given Fig. 4 that results of our calculations are quite well describe the data on capture cross sections at 25 meV, but do not reproduce measurements of Ref. 84 at the energy 65 keV, given by open circle. The reason of this, apparently, is in the absence of taking into account transitions from the resonance $^2S_{1/2}$ and $^4S_{3/2}$ scattering waves with relatively big width to the GS and excited $^{2+4}P_{1/2}$ states with $J^\pi = 1/2^-$.

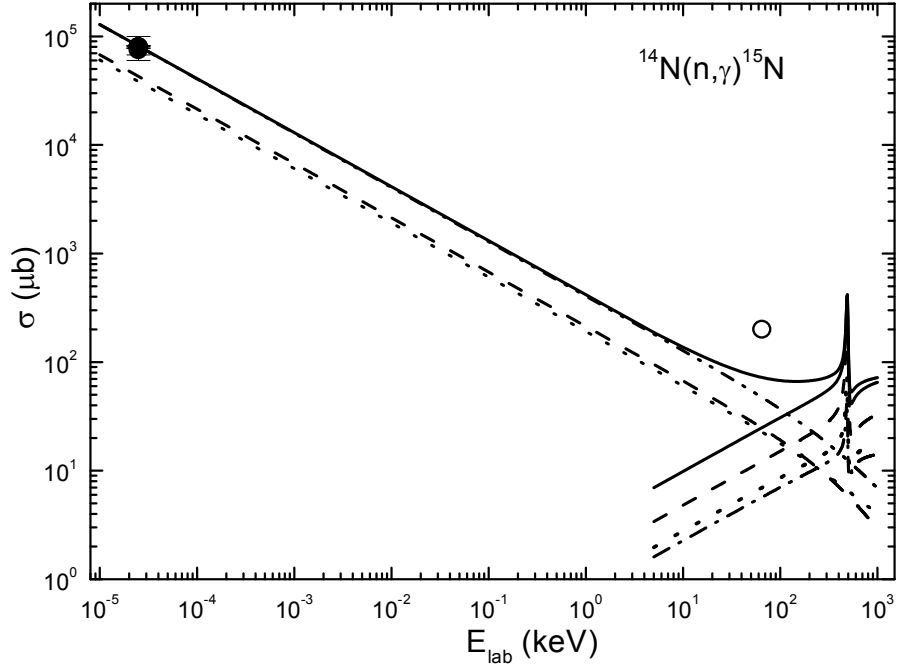

Fig. 4. The total cross sections of the radiative neutron capture on $^{14}$N. Points (•) – represent experiment data from Ref. 77-83, open circles (○) – from Ref. 84. Lines show results of calculations of the total cross sections given in the text.

Furthermore, if, for comparison, we will use the $^2S_{1/2}$ and $^4S_{3/2}$ scattering potentials with zero phases and zero depth, which have not FSs, i.e., do not agree with the given above classification of FSs and ASs according to Young tableaux, then the results of calculations of cross section for potentials of the GS of Eq. (33) and the ES of Eq. (34) lie above all of the experimental data more than by an order.

As it is seen from the obtained results, the description of the total radiative capture cross sections at lowest energies and the BS characteristics, including the AC of $^{15}$N in the n$^{14}$N channel can be well harmonized on the basis of the considered combination of the potentials of Eqs. (31) and (33). In other words, if to fix parameters of the GS potential of $^{15}$N in the n$^{14}$N channel on the basis of the correct description of its characteristics, including AC. Therefore, it is quite possible on the basis of classification of FSs and ASs according to Young tableaux to find such $^{2+4}S_{1/2}$ scattering potentials that allow us to correctly describe not only near-zero elastic scattering phase shifts but also the value of cross sections of the radiative neutron capture on $^{14}$N at the energy of 25 meV. The future measurements of total cross sections for other energies would make it possible to clarify the quality of description of such cross sections in the considered model for potentials with FSs more unambiguously.

Since at the energies from 10 meV to 10 keV the calculated cross section, given in Fig. 4 by the solid line. is almost straight line, it can be approximated by the simple function of energy from Eq. (22). The constant value $A = 406.4817$ μb·keV$^{1/2}$, in this case, was determined by a single point at cross-sections with minimal energy of 10 meV (l.s.). The absolute value from Eq. (23) of relative deviation of the calculated theoretical cross sections and the approximation of this cross section by the given above function from Eq. (7) in the range to 10 keV is less than 0.9%. If to perform an evaluation of the cross section value, for example, at the energy of 1 μeV, result is 12.8 b.

## 5. Radiative neutron capture on $^{15}$N in cluster model

Before proceeding to the total cross-sections analysis of the $^{15}$N(n, γ)$^{16}$N capture reaction, let us note that the classification of orbital states of $^{15}$N according to Young tableaux was qualitatively considered by us in works.[7,85] So for the n$^{15}$N system we have {1} × {4443} → {5443} + {4444}.[46] The first of the obtained tableaux is compatible with the orbital moments $L = 1, 3$ and is forbidden as it has 5 boxes in the first row,[46] and the second one is allowed and compatible with the orbital moments $L = 0$ and 2.[46] Thus, taking into account only the lowest partial waves it may be said that in the potentials of $S$ and $D$ waves the forbidden states are absent, but the $P$ wave contains the forbidden state. The allowed $D$ wave corresponds to the GS of $^{16}$N in the n$^{15}$N system with the binding energy of -2.491 MeV.[86] Since, for the quantum characteristics of $^{15}$N we have values $J^\pi T = 1/2^-1/2$ (Ref. 65) and for $^{16}$N nucleus they are known as $J^\pi T = 2^-1$,[86] then the GS of $^{16}$N in the n$^{15}$N channel may be the mixture of $^1D_2$ and $^3D_2$ states in $^{2S+1}D_J$ notations.

Since we have not the total tables of the Young tableaux product for the particle systems with A > 8,[47] which were used by us earlier for such calculations,[6,7,9-11,32,33,64,87] therefore the obtained above result should be considered only as a qualitative evaluation of possible orbital symmetries for the bound states of $^{16}$N in the n$^{15}$N channel. At once just on the basis of such classification it was succeeded to explain well enough the available experimental data for $^{13}$C(p, γ)$^{14}$N (Refs. 63 and 88), $^{13}$C(n, γ)$^{14}$C (Refs. 6, 64 and 89). Therefore, here we will use the given above classification of cluster states according to the orbital symmetries, which gives the certain number of the FS and AS in the different partial intercluster interaction potentials. The number of these states defines the number of nodes of the radial WF of relative motion of clusters with certain orbital moment.[6,64]

### 5.1. *Interaction potentials*

For the description of $^{15}$N(n, γ)$^{16}$N total cross-sections we considered, as well as in the previous works of Refs. 6, 7, 9-11, 32, 33, 63, 64, 85 and 87-89, the $E1$ transition from the nonresonance $^3P_2$ scattering wave with zero phases at energies up to 1.0 MeV to the triplet part of the $^3D_2$ wave function of the bound ground state of $^{16}$N in the n$^{15}$N channel. In addition, the $E1$ transition is possible from the resonance $P_1$ scattering wave at 0.921 MeV[86] which is the mixture of the $^3P_1$ triplet and the $^1P_1$ singlet states, to the $^3D_2$ triplet and the $^1D_2$ singlet parts of the ground state wave function

$$\sigma_0(E1) = \sigma(E1, {}^3P_2 \to {}^3D_2) + [\sigma(E1, {}^3P_1 \to {}^3D_2) + \sigma(E1, {}^1P_1 \to {}^1D_2)]/2 . \quad (38)$$

The averaging over transitions from the mixed by spins $P_1$ scattering wave to the mixed by spins $D_2$ bound GS of $^{16}$N in the n$^{15}$N channel, because this is one transition from the resonance scattering state to the bound GS of the nucleus, but not two different $E1$ processes. The using model allows us to obtain only mixed by spin WF of n$^{15}$N states. In these calculations the exact value of neutron mass[41] with $^{15}$N mass equal to 15.000108 amu[58] was assigned. For performing the calculations of the radiative capture total cross sections the nuclear part of intercluster n$^{15}$N interaction potential as usual is presented in the Gaussian form of Eq. (8).[6,7,64]

Let us notice that for the potential of resonance $^{1+3}P_1$ scattering waves at 0.921 MeV[86] with one FS it is not possible to obtain the potential which will reproduce correctly the width of the resonance. For example, the potential with parameters $V_{P1} = 7687.40$ MeV, $\alpha_{P1} = 10.0$ fm$^{-2}$ leads to the resonance width equal to 138 keV (c.m.) at the energy 0.921 MeV. That is approximately in 10 times greater than experimental value of 14 keV (l.s.) given in table 16.10 of Ref. 86. Although according to the data presented in table 16.5 of the same review of Ref. 86 this width is equal to 15(5) keV (c.m.). Calculation of the $P_1$ phase shift of the n$^{15}$N elastic scattering with this potential at energies from 0.2 MeV to 1.25 MeV shows that it has the resonance form represented in Fig. 5 by the solid curve. In order that this potential may correctly reproduce the width of the resonance it would be necessary to increase greatly α parameter, i.e., to decrease the width of the potential. For example, the parameters 15385.47 MeV and 20.0 fm$^{-2}$ lead to the width of the resonance about 98 keV (c.m.). This potential phase shift is shown in Fig. 5 by the dotted line. The parameters of much more narrow potential with one FS are equal to

$$V_{P1} = 30781.774 \text{ MeV}, \quad \alpha_{P1} = 40.0 \text{ fm}^{-2}.  \quad (39)$$

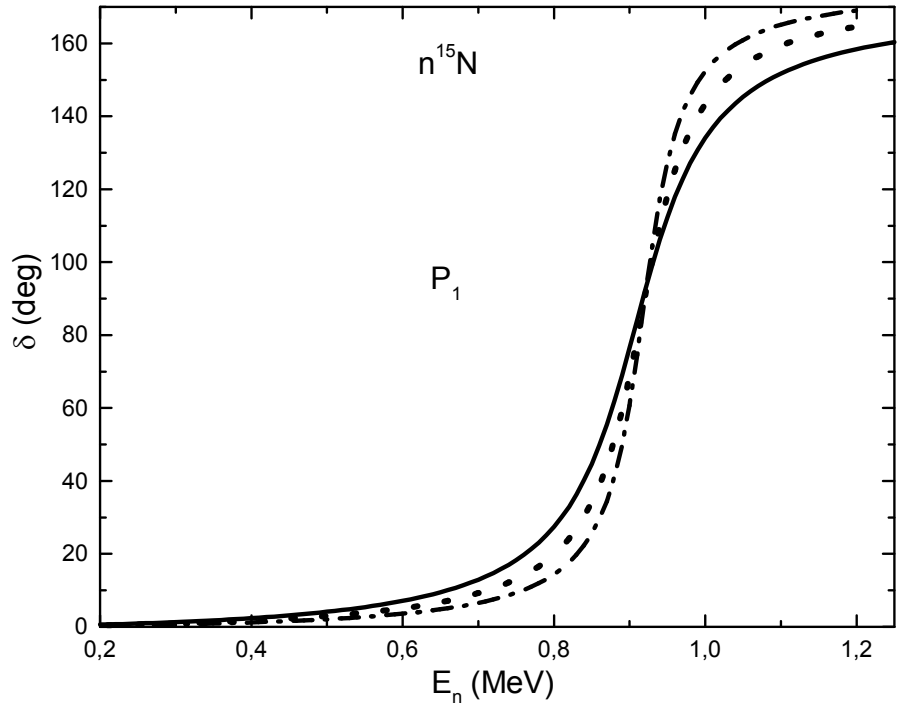

Fig. 5. The phase shifts of the elastic n$^{15}$N scattering in the $P_1$ wave. The lines are obtained for different potentials described in the text.

They lead to the resonance width of 70 keV (c.m.) and its phase is represented in Fig. 5 by the dashed line. Relative accuracy of the $P_1$ scattering phase shift computing in these calculations is approximately $\pm 10^{-3}$, and this potential leads to the phase shift value of 90.0(1)° at the resonance energy of 921 keV.

As one can see from these calculations even for very approximate description of the resonance width in the elastic n$^{15}$N scattering the $P_1$ potential is obtained with absolutely exceptional depth and almost zero width. So for the independent checking of these results other program for the phases calculation was used. This program was independently developed by the outside producer and it is based on other computing methods for the WF calculation and rather different ways of matching of the WF with asymptotic constant. On the basis of it the scattering phase shift of 89.9° was obtained for the potential of Eq. (39) at the resonance energy of 921 keV (l.s.) and specifying of 500 thousand steps of the wave function calculation and matching of it with asymptotic at 30 fm. This result differs from the other one given above only on 0.1°, i.e., on the phase estimation error evaluation specified in our program with automatic selection of step and the wave function matching radius.

Here it is necessary to notice that the potential with the FS at the known energy of resonance level in $^{16}$N spectra and its width are constructed absolutely unambiguously. It is not possible to find other parameters $V_0$ and α which would be able to reproduce correctly the resonance level energy and its width, if the number of FS is given, which in this case is equal to 1. Such potential depth unambiguously locates a position of resonance, i.e., the energy of the resonance level, and its width gives the definite width of this resonance state. However, we didn't succeed to find some physical interpretation of so small width and huge depth of this potential. The total mass of 16 nucleons in $^{16}$N is approximately equal to 15 GeV, and the depth of potential of Eq. (39) takes on a value 30 GeV. Thus, used here potentials should be considered as an attempt of rough approximation within the potential approach (i.e., on the basis of interaction potentials taking into account a resonance behavior of scattering phases of two nuclear particles with given masses) of experimentally observed width of the considered level at energy 921 keV which is present in the elastic n$^{15}$N scattering (see table 16.10 from review of Ref. 86).

For the potentials of non-resonance $^3P_2$ and $^3P_0$ waves with one FS we have used the parameters' values based on the assumption that in the considered range of energies, i.e., up to 1.0 MeV, the potentials phase shifts are equal to zero, as the resonance levels with $J = 2^+$ and $0^+$ (see table 16.10, Ref. 86) are not observed in $^{16}$N spectra in the n$^{15}$N channel. Particularly, it has been obtained that

$$V_{P2} = 500.0 \text{ MeV}, \quad \alpha_{P2} = 1.0 \text{ fm}^{-2}. \tag{40}$$

Calculation of the *P* phase shifts with such potential at energy up to 1.0 MeV leads to their values less than 0.1°.

Furthermore, we will construct the potential of the $^{1+3}D_2$ ground state without the bound FS, which must correctly reproduce the binding energy of $^{16}$N ground state with $J^{\pi}T = 2^-1$ in the n$^{15}$N channel at -2.491 MeV.[86] Also this potential must reasonably describe a mean square radius of $^{16}$N, experimental value of which evidently must not exceed a lot the radius of $^{16}$O equal to 2.710(15) fm.[86] Note, that the experimental radius of $^{15}$N is equal to 2.612(9) fm.[65] In these calculations we use a zero charge radius of neutron the mass radius of which is equal to proton radius 0.8775(51) fm.[41] Consequently, the following parameters for the GS potential of $^{16}$N the in n$^{15}$N channel were obtained

$$V_{G.S.} = 49.5356532 \text{ MeV}, \quad \alpha_{G.S.} = 0.07 \text{ fm}^{-2}. \tag{41}$$

The potential leads to the binding energy of -2.49100003 MeV at the FDM accuracy of $10^{-8}$ MeV, the mean square charge radius of 2.63 fm and the mass radius of 2.76 fm. For the AC defined in Ref. 40 and written in the non-dimensional form of Eq. (7) the value of 0.96(1) was obtained over the range of 6–19 fm. Error of the constant is defined by its averaging over the mentioned above distance range. In Ref. 27 this AC has a value of 0.85 fm$^{-1/2}$, that after the recalculation to the non-dimensional quantity at $\sqrt{2k_0}$ = 0.821 leads to the value of 1.04. This value differs from the other one obtained above only on 10%. The recalculation of the AC is needed because in Ref. 27 other definition of the AC was used

$$\chi_L(R) = CW_{-\eta L+1/2}(2k_0 R), \qquad (42)$$

that differs from the used here definition on a factor $\sqrt{2k_0}$. Let us note that in Ref. 27 the AC also were obtained for the first three excited states of $^{16}$N in the n$^{15}$N channel with $J$ = 0$^-$, 3$^-$ and 1$^-$, their values are equal to 1.10, 0.29 and 1.08 fm$^{-1/2}$ that after recalculation to the non-dimensional form gives: 1.34, 0.35 and 1.32. Furthermore, the potentials of these three bound ESs are constructed so that to describe correctly AC data, which were obtained in Ref. 27 and given above.

For additional control of calculation of the ground state energy the variational method[35,66] was used. This method for the grid dimension $N$ = 10 and independent variation of parameters for potential of Eq. (41) let to obtain the energy of -2.49100001 MeV. The value of the AC is equal to 0.96(1) over the range of 9–24 fm, and the charge radius doesn't differ from other one obtained above within the FDM. As for the basis dimension increasing the variational energy decreases and gives the upper limit of true binding energy,[6,64] and for the stride parameter decreasing and steps number increasing the finite-difference energy increases,[6,7,35,64,66] so the real binding energy in such potential can take on the average value -2.49100002(1) MeV. Thus, the accuracy of calculation of two-body binding energy of $^{16}$N by using two methods (the FDM and the variational method) and two different and independent computer programs is equal to ±10$^{-8}$ MeV = ±10 meV. This value is agree with the primordially given in the FDM accuracy in case of calculation of two-cluster system binding energy.

Besides the GS potential, the potentials of three ES at energies of 0.12042 MeV with $J$ = 0$^-$, 0.29822 MeV with $J$ = 3$^-$ and 0.39727 MeV with $J$ = 1$^-$ (Ref. 86) relatively the GS of $^{16}$N or -2.37058, -2.19278 and -2.09373 MeV relative to the threshold of the n$^{15}$N channel. The $^1S_0$, $^3D_3$ and $^3S_1$ levels of $^{16}$N under its consideration in the n$^{15}$N cluster channel could be matched to these bound states in the n$^{15}$N channel. In consequence of accounting of these ES, the next $E$1 transitions are considered additionally: from the nonresonance $^3P_2$ and $^3P_0$ waves and from the resonance, mixed by spin $^{1+3}P_1$ scattering state

$$\sigma_{ex}(E1) = \sigma(E1, {}^3P_0 \to {}^3S_1) + \sigma(E1, {}^3P_1 \to {}^3S_1) + \sigma(E1, {}^3P_2 \to {}^3S_1) + \\ + \sigma(E1, {}^3P_2 \to {}^3D_3) + \sigma(E1, {}^1P_1 \to {}^1S_0). \qquad (43)$$

The parameters of the $^1S_0$ potential without FS in the n$^{15}$N channel were obtained for the first of these ES:

$$V_{S0} = 54.454312 \text{ MeV}, \quad \alpha_{S0} = 0.6 \text{ fm}^{-2}. \qquad (44)$$

The potential leads to the biding energy of -2.370580 MeV at the FDM accuracy of $10^{-6}$ MeV, the mean square charge radius of 2.62 fm and the mass radius of 2.63 fm, and for the AC value of 1.35(1) in the interval 3–22 fm, which is in a good agreement with the given above results.[27]

The parameters of the $^3D_3$ potential without FS in the n$^{15}$N channel were obtained for the second ES:

$$V_{D3} = 126.14123 \text{ MeV}, \quad \alpha_{D3} = 0.2 \text{ fm}^{-2}. \qquad (45)$$

The potential gives the binding energy of -2.192780 MeV at the FDM accuracy of $10^{-6}$ MeV, the mean square charge radius of 2.62 fm and the mass radius of 2.64 fm. The AC value is equal to 0.32(1) in the interval of 5–22 fm, which is in a good agreement with the given above results.[27]

The next parameters of the $^3S_1$ potential without FS in the n$^{15}$N channel were obtained for the third ES:

$$V_{S1} = 53.170538 \text{ MeV}, \quad \alpha_{S1} = 0.6 \text{ fm}^{-2}. \qquad (46)$$

The potential leads to the biding energy of -2.093730 MeV at the FDM accuracy of $10^{-6}$ MeV, the mean square charge radius of 2.62 fm and the mass radius of 2.64 fm, and for the AC value of 1.33(1) in the interval of 3–23 fm, which, in practice, does not differ from the results of work.[27]

### 5.2. *Total cross sections of the neutron capture on $^{15}$N*

Going to the immediate consideration of the results for the mentioned above $E1$ transitions to the ground state and the first three excited states of $^{16}$N let us note, that we succeeded to find the experimental data[49,51,90,91] for the total cross section for the $^{15}$N(n, γ)$^{16}$N process only at three energy values 25, 152 and 370 keV,[92] these results are presented in Figs. 6a–6c by the black dots. The results of our calculation of the total cross section of the $E1$ capture process ($^3P_2 \to\ ^3D_2$) to the ground state are shown in Fig. 6a by the dotted line, the cross sections for [σ($E1$, $^3P_1 \to\ ^3D_2$) + σ($E1$, $^1P_1 \to\ ^1D_2$)]/2 transitions also to the ground state are presented by dashed line, and their sum is given by solid line. The given above potentials of Eqs. (39)–(41) were used in these calculations.

Then it may be noted, that if the parameters of the $P_1$ resonance potential are fixed over the phase resonance rather unambiguously, and for the bound state they are definitely selected on the basis of the description of its characteristics, so the parameters of the $^3P_2$-potential with the FS of Eq. (40) leading to zero phases may take on other values. However, if one use, for example, more narrow potential with parameters 1000 MeV and 2.0 fm$^{-2}$ which also leads to the near-zero phases, the results of the cross section calculations for the transition to the ground state differ on the value about 1%. This result demonstrates the weak influence of such potential geometry on the capture total cross sections. Here only the near-zero value of scattering phase is important.

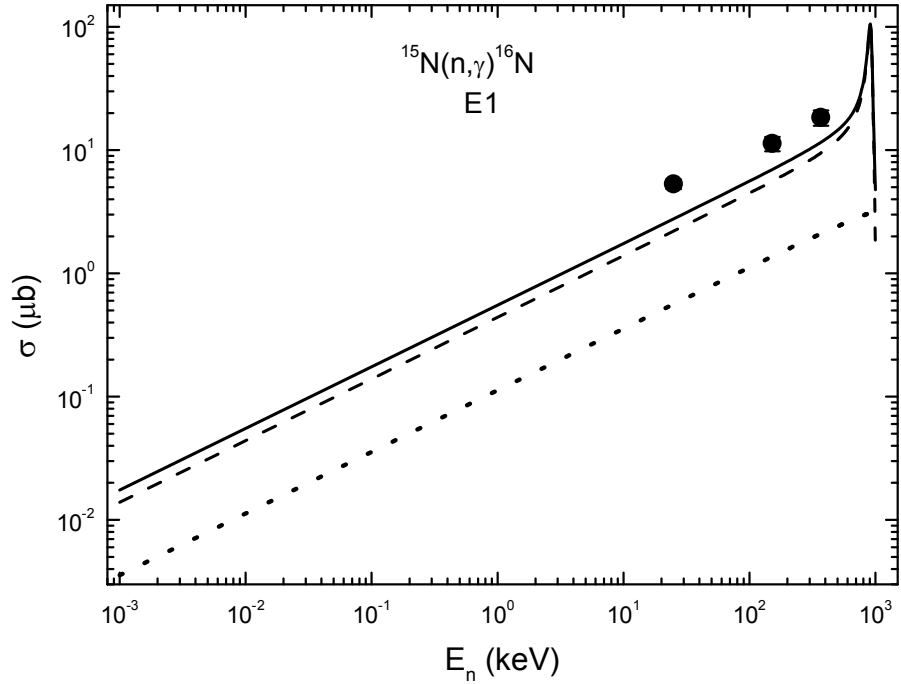

Fig. 6a. The total cross sections of the radiative neutron capture on $^{15}$N. Experimental data: ● – Ref. 92. Lines – calculations of the total cross sections for the transitions to the GS.

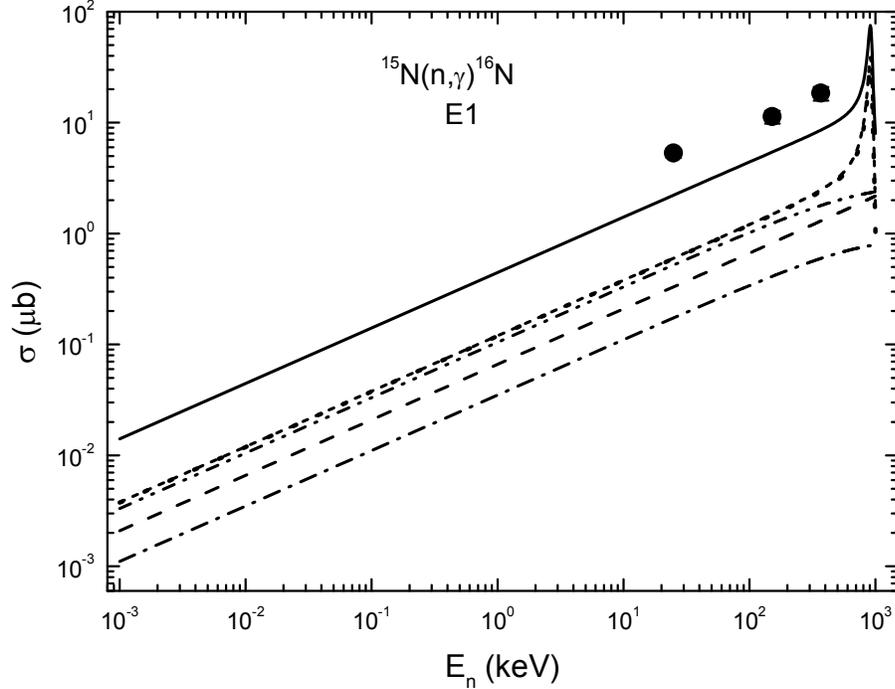

Fig. 6b. The total cross sections of the radiative neutron capture on $^{15}$N. Experimental data: ● – Ref. 92. Lines – calculations of the total cross sections for the transitions to the ESs.

Furthermore, the results of calculations of total cross sections of five $E1$ transitions to three ESs are shown in Fig. 6b in the following way: the $\sigma(E1, {}^3P_0 \to {}^3S_1)$ transition is shown by the dot-dashed line, the $\sigma(E1, {}^3P_1 \to {}^3S_1)$ process is shown by the dotted line, the $\sigma(E1, {}^3P_2 \to {}^3S_1)$ is shown by the dot-dot-dashed line, the $\sigma(E1, {}^3P_2 \to {}^3D_3)$ is shown by the usual dashed line, and the $\sigma(E1, {}^1P_1 \to {}^1S_0)$ is shown by the short dashed line, which is practically matching with the dotted line for the $\sigma(E1, {}^3P_1 \to {}^3S_1)$ process. The solid line shows the summed total cross section over all of these $E1$ transitions to all three ESs. The potentials of Eqs. (39) and (40), and also of Eqs. (44)–(46) were used for these calculations.

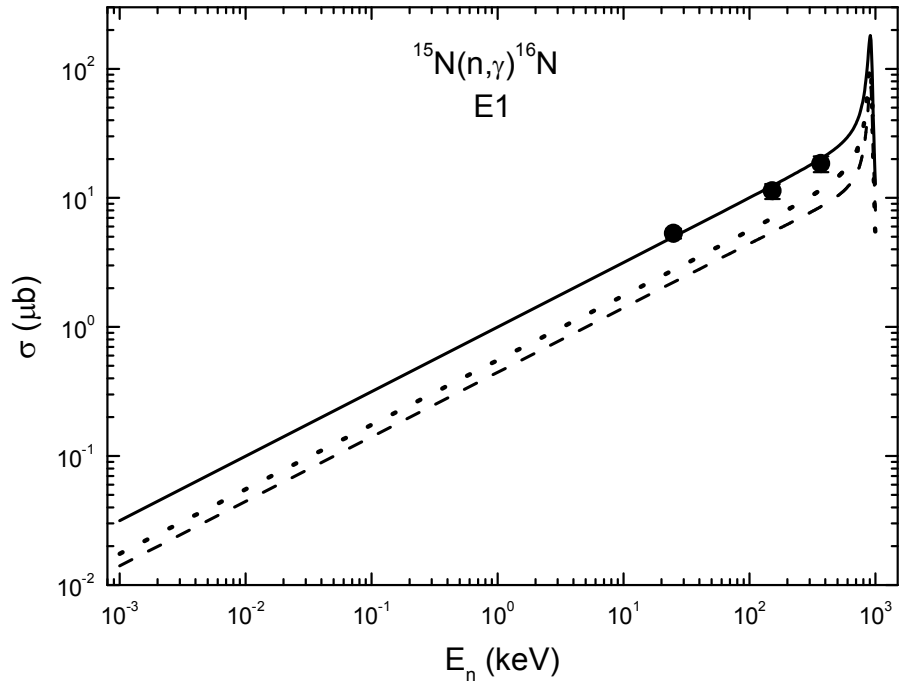

Fig. 6c. The total cross sections of the radiative neutron capture on $^{15}$N. Experimental data: ● – Ref. 92. Lines – calculations of the total cross sections for the transitions to the GS and ES in the energy range from 1 eV to 1 MeV.

Finally, Fig. 6c shows the total cross sections for the transitions to the GS (dotted line – this cross section is shown in Fig. 6a by the solid line), and to the first three ES – this cross section is shown by the dashed line, which corresponds to the solid line in Fig. 6b. The sum of all considered processes is shown in Fig. 6c by the solid line, which describes the available experimental data[92] quite well. We need to note that due to the large width of the resonance of the phase shift, which gives the $P_1$ potential of Eq. (39), the width of the resonance of total cross sections in these calculations is also substantially overestimated. This leads to slightly overestimated value of the cross section at the energy of 370 keV,[92] located near the resonance.

Thus, the intercluster potentials of the different bound states, constructed on the basis of quite obvious requirements for description of the binding energy, the mean square radii of $^{16}$N, the AC values in the n$^{15}$N channel, are allowed correctly reproduce the existent experimental data for total cross sections of the radiative neutron capture on $^{15}$N at low energies.[92] Besides, all used n$^{15}$N potentials were constructed on the basis of the given above classification of FSs and ASs according to Young tableaux.

However, it is difficult to draw certain conclusions keeping in hands only three experimental data of total cross sections at 25–370 keV.[92] Therefore, it is desirable in future to do the measurements of such cross sections in the energy range from 1–10 keV to 1.0–1.2 MeV. These data should acceptably estimate the width of the resonance of this reaction at 921 keV[86] and the value of the cross section at the resonance energy.

This allows one to compare it with the results of the given calculations, which predict the cross section value in the resonance at 180 μb.

Now let us notice, as the calculated cross section is practically the straight line at the lowest energies from 1 eV to 10 keV (see the solid line in Fig. 6c), then it may be approximated by the simple function of the form

$$\sigma_{ap} = 0.9968\sqrt{E_n} \ . \tag{47}$$

The given constant's value 0.9968 μb·keV$^{-1/2}$ has been defined over one point in the cross-sections at the minimal energy equal to 1 eV. Then it was found that the absolute value of the relative departure of the calculated theoretical cross-section and approximation of this cross-section by the given above function of Eq. (23) at energies less than 10 keV is about 0.1%. If it is assumed, that this form of energy dependence of the total cross section will be also conserved at lower energies, then one may give estimate of the cross section value which, for example, at energy 1 meV ($10^{-3}$ eV) is equal to 9.97·$10^{-4}$ μb.

## 6. Radiative neutron capture on $^{16}$O in cluster model

### 6.1. *Classification of the orbital states*

Going to the analysis of the total cross sections of the $^{16}$O(n, γ)$^{17}$O capture, let's first consider the classification of the orbital states of the n$^{16}$O system according to Young tableaux. The GS of $^{16}$O corresponds to the Young tableau {4444}.[5,9,93] Recall that the possible orbital Young tableaux in the $N = n_1 + n_2$ particles system can be defined as the direct exterior product of the orbital tableaux of each subsystem, that gives {1} × {4444} → {5444} + {44441} for the n$^{16}$O system.[46] The first of these tableaux is compatible with the orbital angular moment $L = 0$ and it is forbidden, because five nucleons can not be in the *s*-shell, and the second tableau is allowed and compatible with the orbital angular moment equal to 1.[46]

Hence, in the potential of $^2S_{1/2}$ wave that corresponds to the first excited state of $^{17}$O in the n$^{16}$O channel and scattering states of these clusters, there is a forbidden state, and the $^3P$ scattering wave does not contain the forbidden state, but the allowed BS with {44441} tableau can be in both spectra – continuous and discrete. The GS of $^{17}$O in the n$^{16}$O channel, which has the energy of -4.1436 MeV,[94] refers to the $^2D_{5/2}$ wave and also does not contain FSs. However, since we do not have the complete tables of products of Young tableaux for the system with the number of particles greater than eight[47] previously used for such calculations,[9-11,32,87] then obtained above results should be considered only as a qualitative assessment of possible orbital symmetries in the bound states of $^{17}$O in the n$^{16}$O channel.

### 6.2. *Phases and potentials*

To perform the calculation of radiative capture within the MPCM it is need to know the potentials of the n$^{16}$O elastic scattering in $^2S_{1/2}$, $^2P_{1/2}$, $^2P_{3/2}$, $^2D_{3/2}$ and $^2D_{5/2}$ waves and the interaction of the ground $^2D_{5/2}$ and the $^2S_{1/2}$ first excited states of $^{17}$O in the

n$^{16}$O channel. The experimental data for the total cross sections of radiative capture, obtained in Refs. 95 and 96], are presented for the transition just to these bound states.

As already mentioned, the scattering potentials are constructed on the basis of the elastic scattering phases obtained at the energy $E > 1.1$ MeV in Refs. 97 And 98]. For the energy range 0.2–0.7 MeV there are the results of the phase analysis[99] based on the measurements of differential cross sections for n$^{16}$O elastic scattering[100] in the resonance region at 0.433 MeV.[94] Later, new experimental data[101] on the excitation function at energies from 0.5 MeV to 6.2 MeV were given in the EXFOR database.[102,103] These data[101] have not been used to date in the phase shift analysis in the $^2D_{3/2}$ resonance region at the energy of 1.0 MeV.[94]

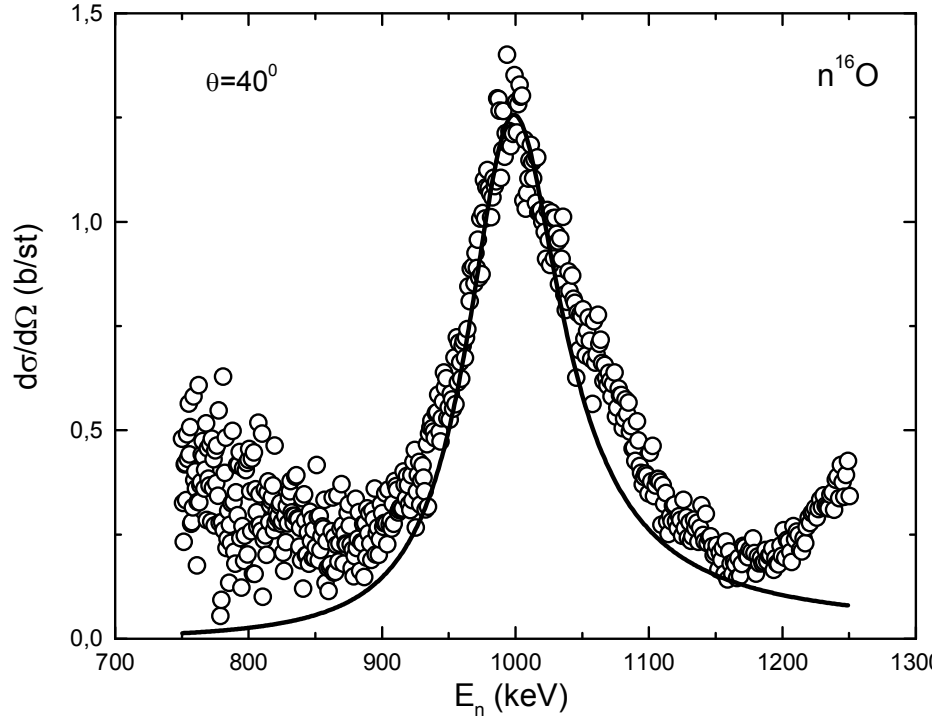

Fig. 7. The excitation functions of the n$^{16}$O elastic scattering in the $^2D_{3/2}$ resonance region at 1.0 MeV. The data from Ref. 101 are shown by the open circles. Solid line is the calculation of the cross sections with the potential given in the text.

Here, the data from Ref. 101 were used to perform the phase shift analysis and for extraction the form of phase in the $^2D_{3/2}$ scattering wave. The used excitation functions at 40° (l.s.)[101] are shown in Fig. 7 at the energy range from 0.75 MeV to 1.25 MeV (l.s.) by the open circles. The experimental errors which in some points reach 25% are not presented in Fig. 7, because they overload a lot the figure. This is caused by that in our analysis more than 500 points were used for the cross sections at the different energies from the excitation functions.[101] First it should be noted that below 0.7–0.8 MeV the ambiguity of the data from Ref. 101 increases dramatically, however, to extract $^2D_{3/2}$ scattering phase shift is sufficient to consider the energy region shown in Fig. 7, which has a relatively small ambiguities and these data may be well used for carrying out the phase analysis. Earlier, we have performed the phase analysis in the similar systems: n$^{12}$C (Refs. 7 and 104); p$^{12}$C Ref. 105; p$^6$Li Ref. 106; and p$^{13}$C Refs. 88 and 107, mainly at the astrophysical energies.

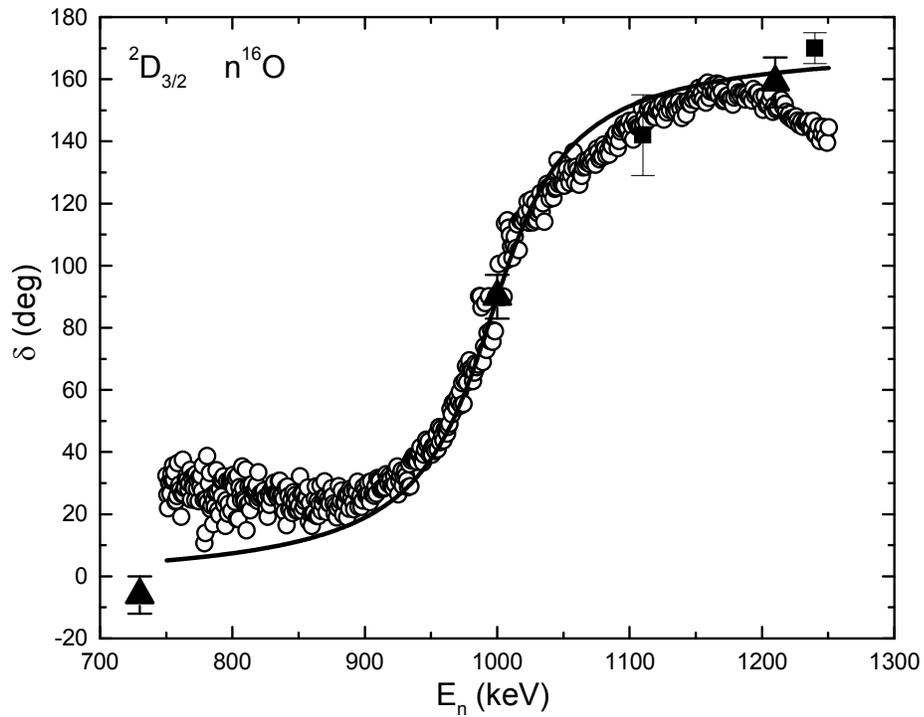

Fig. 8a. The $^2D_{3/2}$ phase shift of the n$^{16}$O elastic scattering at low energies. Open circles (○) are the results of our phase analysis carried out on the basis of the data from Ref. 101; black squares (■) show the results of phase analysis of Ref. 97; triangles (▲) correspond to the results of the analysis of Ref. 99; solid curve is the calculation of phase shift with the potential given in the text.

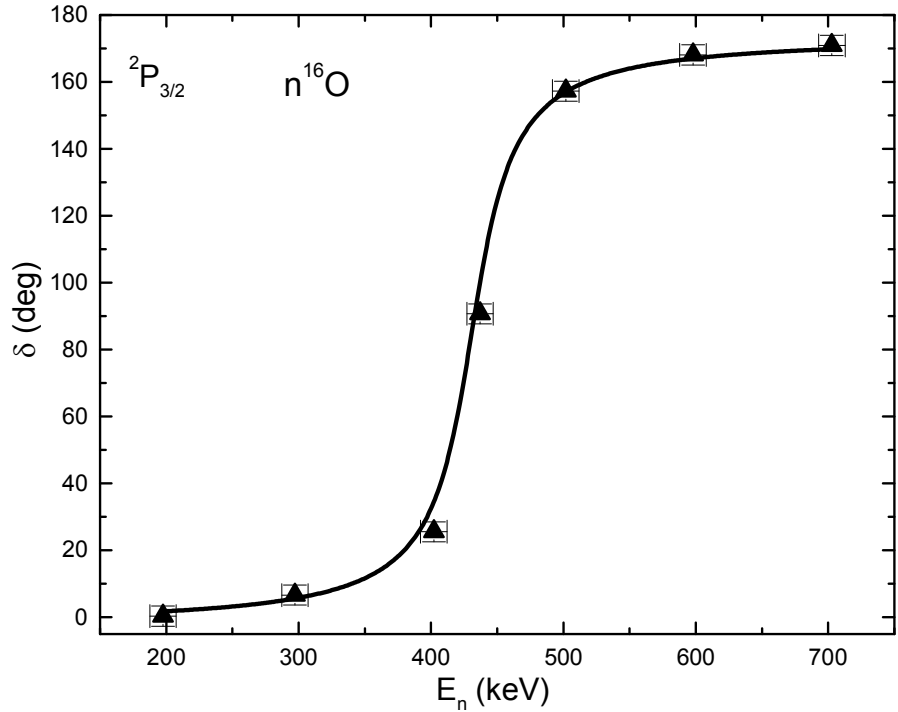

Fig. 8b. The $^2P_{3/2}$ phase shift of the n$^{16}$O elastic scattering at low energies. Triangles (▲) show the results of the phase shift analysis of Ref. 99. The solid curve corresponds to the calculation of phase shift with the potential given in the text.

The details of the used method of searching the phase shifts in the elastic scattering of particles with spin $1/2^+0$ are presented in Ref. 7, the main expressions are given in Refs. 7 and 8, and the results of the present analysis of the n$^{16}$O elastic scattering in the energy range from 0.75 MeV to 1.25 MeV are shown by the open circles in Fig. 8a. Black squares in Fig. 8a shows the results of the phase analysis of Ref. 97 obtained at the energy $E > 1.1$ MeV, and the triangles are the results of the analysis of Ref. 99. The quantity $\chi^2$ has a mean value of $4.7 \cdot 10^{-3}$ with a maximum value of the partial quantity $\chi^2_i = 0.6$ at the energy of 999.5 keV, since only one value of the cross sections is considered for each energy value. To describe the cross sections in the excitation functions,[101] at least at energies up to 1.2–1.25 MeV, is not required to take into account the $^2S_{1/2}$ scattering phase, because its presence does not change the magnitude of $\chi^2$, i.e., its value can be assumed to be equal to zero.

For a description of the $^2D_{3/2}$ phase shift obtained from the phase analysis, one can use the potential of the form of Eq. (8) without the FS with the parameters:

$$V_D = 95.797 \text{ MeV, and } \alpha_D = 0.17 \text{ fm}^{-2}, \tag{48}$$

which leads to the resonance energy of 1000 keV at the phase shift of 90.0 (0.1)° with the level width of 88 keV (l.s.) or 83 keV (c.m.). At the same time, in Table 17.17 of Ref. 94 the width is equal to 102 keV (l.s.) or 96 keV (c.m.) at the level energy of 1000±2 keV (l.s.). The energy dependence of the $^2D_{3/2}$ phase shift of potential of Eq. (48) is shown in Fig. 8a by a solid curve. Such a potential describes well the behavior of the scattering phase shift in the resonance region and is consistent with the previous scattering phase shift extractions.[97,99] The shape of the cross sections in the excitation functions, calculated with the $^2D_{3/2}$ phase shift of potential of Eq. (48) at zero values of other phases is shown in Fig. 7 by the solid curve. To estimate the width of the resonance was used expression of Eq. (16).

Next, we will consider the total radiative capture cross section with the $E1$ transitions from the $^2P_{3/2}$ resonance in n$^{16}$O scattering at 433 keV to the ground and first excited states of $^{17}$O. To build the $^2P_{3/2}$ scattering potential the data on the position and width of this level from the review of Ref. 94 (see Table 17.17) and the results of the phase analysis[99] shown by triangles in Fig. 8b have been used. As a result, is found that to describe the resonant $^2P_{3/2}$ scattering phase shift at 433(2) keV (l.s.) with the width of 45 keV (c.m.) or 48 keV (l.s.)[94,100] the potential without the FS with the following parameters is needed:

$$V_P = 1583.545 \text{ MeV, and } \alpha_P = 6.0 \text{ fm}^{-2}, \tag{49}$$

which leads to the level width equal to 44 keV (c.m.) or 47 keV (l.s.) at the resonance of 433 keV (l.s.), i.e., its phase at this energy is equal to 90.0(0.2)°, and the total dependence of the phase on the energy in the resonance region is shown in Fig. 8b by the solid curve.

The $V_0 = 0$ MeV, i.e. zero phase shifts, were used for the potentials of the non-resonance $^2P_{1/2}$ and $^2D_{5/2}$ scattering waves, because in spectra of $^{17}$O below 1.0–1.5 MeV the levels with $J = 1/2^-$ and $5/2^+$ are not observed, and they have no FSs. Next, one should note that the potential is built unambiguously on the basis of known energy of the resonance level in spectra of $^{17}$O (see Table 17.17 Ref. 94) and its width, i.e., it is impossible to find other parameters $V$ and $\alpha$, which can properly reproduce the resonance energy of the level and its width if the number of the FSs is given. In this case the number of the FSs is equal to zero. The depth of this potential unambiguously determines the position of the resonance, i.e., the resonance energy of the level, and the level width sets a certain width of this resonance state.

Next to perform the calculations of the radiative capture within the MPCM the potentials of n$^{16}$O clusters interaction in the bound states are needed. The electromagnetic transitions to the ground bound state of $^{17}$O in the n$^{16}$O channel[94] with $J^\pi, T = 5/2^+, 1/2$ at the energy -4.1426 MeV and to the first excited state of this nucleus with $J^\pi = 1/2^+$ at -3.2729 MeV will be considered. The width of such potentials was fixed on the basis of the correct description of the binding energy and the charge radius of $^{17}$O that is equal to 2.710(15) fm,[94] and then the comparison of the asymptotic constants of the n$^{16}$O channel with other data was performed.

As a result, for the $^2D_{5/2}$ potential of the GS of $^{17}$O in the n$^{16}$O channel without FS the following parameters were found:

$$V_{D0} = 102.2656782 \text{ MeV, and } \alpha_{D0} = 0.15 \text{ fm}^{-2}. \quad (50)$$

These parameters allow to get the binding energy of -4.1436000 MeV with the accuracy of 10$^{-7}$ MeV, the charge radius is equal to 2.71 fm, the mass radius of 2.73 fm and the asymptotic constant on the distance range 6–16 fm equal to $C_W = 0.75(1)$, defined by using the given above expression.[40] The charge radius of the neutron was equal to zero, its mass-radius was assumed to be equal to the proton radius of 0.8775(51) fm,[31] and the value of the charge radius of $^{16}$O was equal to 2.710(15) fm.[94] Methods of calculation of the energy, the wave functions and the radii are given in Refs. 6, 7, 9, and 10. In Ref. 27 for the AC of the ground state the value of 0.9 fm$^{-1/2}$ was obtained, which after recalculation with the factor $\sqrt{2k} = 0.933$ to a dimensionless quantity gives the value 0.96. This recalculation is required, since in these works the different definition of the AC was used, which is different from ours by a factor $\sqrt{2k}$. In Ref. 40 for this constant in dimensionless form the value 0.77(8) is given, which practically coincides with the value obtained above.

For the $^2S_{1/2}$ potential of the first excited state of $^{17}$O in the n$^{16}$O channel with one FS the following parameters were found

$$V_{S1} = 81.746753 \text{ MeV, and } \alpha_{S1} = 0.15 \text{ fm}^{-2}, \quad (51)$$

parameters of Eq. (51) lead to the binding energy of -3.2729000 MeV relative to the threshold of the n$^{16}$O channel of $^{17}$O with the accuracy of 10$^{-7}$ MeV, the charge radius of 2.71 fm, the mass radius of 2.80 fm and AC on the distance range 6–17 fm is equal to $C_W = 3.09(1)$. The value 3.01 fm$^{-1/2}$ for the AC for this level was obtained in Ref. 27, that after the recalculation with $\sqrt{2k} = 0.88$ gives 3.42. As is seen, in this case the asymptotic constant value is in reasonable agreement with the other results.

To control the accuracy of the BS energy calculation a variational method[35] was used. This method for the GS on the grid with the dimension $N = 10$ and the independent variation of the parameters for the potential of Eq. (50) allowed to obtain the energy of -4.1435998 MeV. The charge radius and the asymptotic constant in the interval of 6–16 fm do not differ from the values obtained in the FDM calculations. Since the variational energy is decreasing with the basis dimension increasing and gives the upper limit of the true binding energy, and the finite-difference energy is increasing with decreasing of the step size and increasing of the number of steps,[6,7,9,12,35] so the real binding energy in this potential can take on the average value of -4.1435999(1) MeV. Thus, the accuracy of the binding energy calculation by two methods and two different computer programs is at the level of ±0.1 eV and in full compliance with the defined in the FDM program error of the binding energy search $10^{-7}$ MeV.

For the energy of the first excited state on the grid with the dimension $N = 10$ and the independent variation of the parameters for the potential of Eq. (51) the energy of -3.2728998 MeV was obtained. The charge radius and the AC in the interval of 6–20 fm do not differ from the values obtained within the FDM. Here, the real binding energy can take on the average value of -3.2728999(1) MeV, i.e. the accuracy of the energy calculation by two ways and two different computer programs is also equal to ±0.1 eV.

### 6.3. *Total cross sections of the radiative capture*

Earlier, the radiative neutron capture reaction on $^{16}$O was examined basing on the model of direct capture in Ref. 108, where it was shown that it is possible to describe the available experimental data[95,96] in the range of 20–280 keV. Then basing on the folding model[109] was shown that one managed to describe the experimental data[95,96] in the energy range 20–60 keV. Furthermore, on the basis of the GCM, taking into account only the $E1$ transition in Ref. 110,111 the correct description of the total cross section in the energy range from 20 to 280 keV[95,96] have been obtained. Finally in Ref. 112 on the basis of the GCM and the microscopic R-matrix analysis based on the $E1$ and $M1$ processes on the whole one managed to reproduce correctly the experimental total cross sections at the energies from 25 meV Refs. 78, 113, and 114 to 20–280 keV Refs. 95, 96, i.e., to the resonance at 433 keV corresponding to the $^2P_{3/2}$ wave of n$^{16}$O-scattering,[94] and to predict the possible behavior of the cross sections in the range of the $^2D_{3/2}$ resonance. However, the energies near the $^2P_{3/2}$ resonance, at which there is a comparatively new experimental data[115] from 160 to 560 keV, have not been considered till now.

In the present analysis based on the MPCM with the FS[6,7,9-12] by considering the data[115] the total cross sections of the neutron capture reaction on $^{16}$O in the energy range 10 meV–1.3 MeV will be considered. In this case one takes into account the $E1$ transitions:
1. from the resonance $^2P_{3/2}$ scattering wave to the $^2D_{5/2}$ ground bound state;
2. from the $^2P_{1/2}$ and $^2P_{3/2}$ scattering waves to the $^2S_{1/2}$ first excited state which is bound in the n$^{16}$O channel of $^{17}$O.

In addtion, the following M1 processes will be discussed:
3. from the resonance $^2D_{3/2}$ wave to the $^2D_{5/2}$ ground state of $^{17}$O;
4. from the $^2S_{1/2}$ scattering wave to the $^2S_{1/2}$ first excited state.

The total cross sections of the other possible transitions, for example, the $E2$ transition from the resonance $^2D_{3/2}$ wave to the $^2S_{1/2}$ first excited state or the $M1$ transition from the $^2D_{5/2}$ non-resonance scattering wave to the $^2D_{5/2}$ ground state and others, are on 2–3 orders less than ones mentioned above. Separate consideration of the transitions to the GS and the first ES has been made possible by the measurements carried out in Ref. 95,96 in the energy range 20–280 keV. Moreover, the results of measurements[95,96] have shown that the cross sections of the capture to the first excited state is in 3–4 times greater than another one to the GS of $^{17}$O.

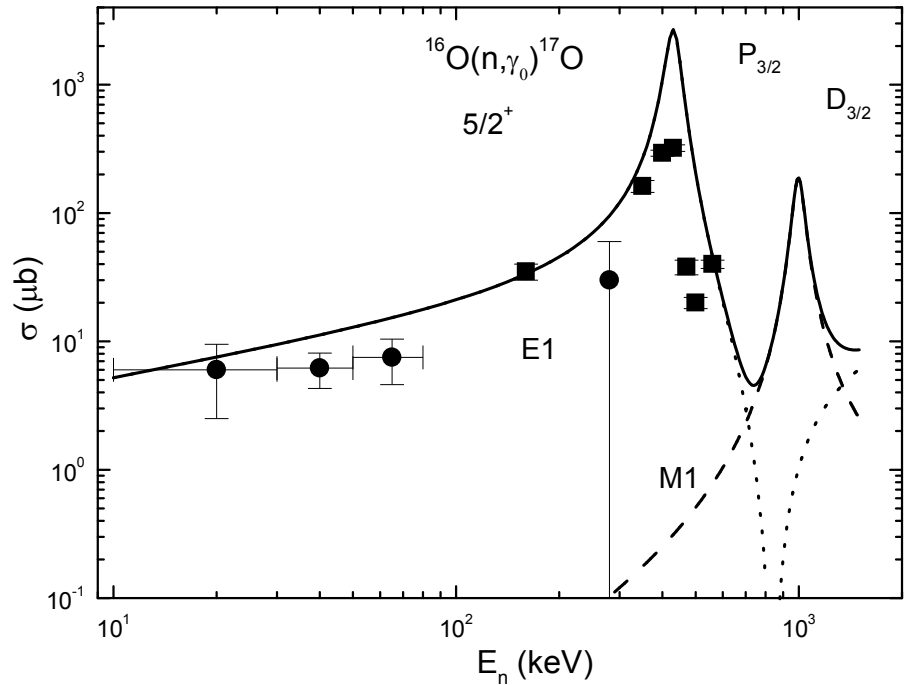

Fig. 9a. The total cross sections of the $E$1 radiative neutron capture on $^{16}$O to the 5/2$^+$ ground state. The experimental data: ● – Refs. 95 and 96, ■ – Ref. 115. Curves correspond to the present calculations with the given in the text potentials.

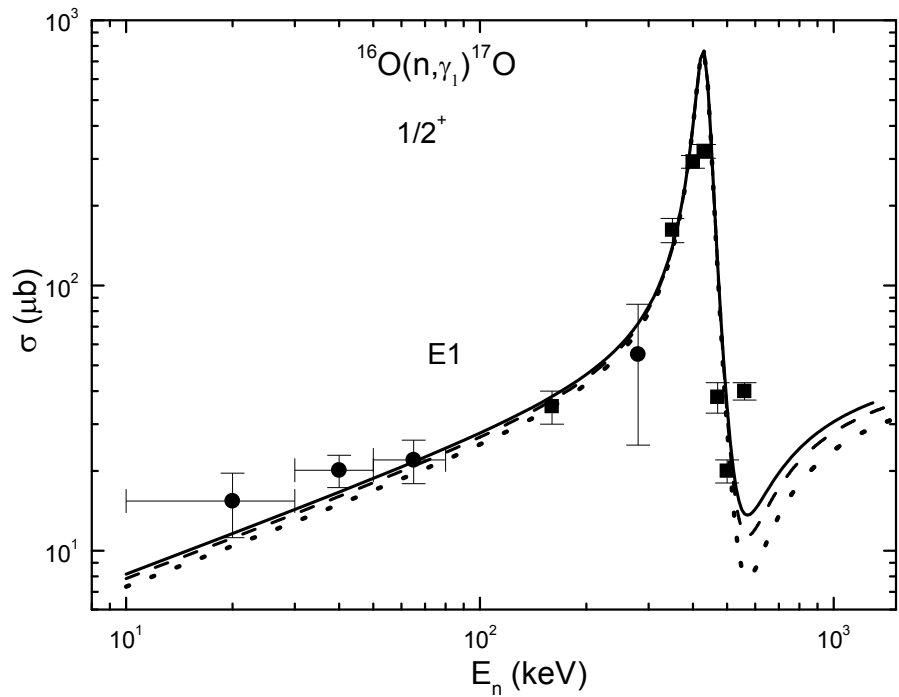

Fig. 9b. The total cross sections of the $E$1 radiative neutron capture on $^{16}$O to the 1/2$^+$ first excited state of $^{17}$O. Experimental data: ● – Refs. 95 and 96, ■ – Ref. 89. Curves are our calculation of the total cross section for the potentials given in the text.

The results of our calculations of the total cross section for the *M*1 and *E*1 transitions to the GS, i.e., the processes 1 and 3 from the given above list, with the potentials of Eqs. (48-50) in comparison with experimental data[95,96,115] are presented in Fig. 9a by the solid curve. The cross sections of *M*1 transition from the $^2D_{3/2}$ scattering wave at energies up to 1.3 MeV, i.e., near the $^2D_{3/2}$ resonance, are represented by the dashed curve (process 3), and for the *E*1 process by the dotted curve (process 1). In Fig. 9a the results of measurements of the total cross sections, performed in the resonance region from 160 to 560 keV in Ref. 89. In Ref. 89, apparently, the summary data for the transitions to the GS and the first excited state are given. From Fig. 9a is seen that our calculations, performed with taking into account the *M*1 and *E*1 processes, reproduce acceptably the results of experimental measurements of the total cross sections of the transition to the GS of $^{17}$O,[95,96] which gradually decreases with the energy decreasing. And it is necessary to note that the potentials of the $^2P_{3/2}$ and $^2D_{3/2}$ scattering waves and the $^2D_{5/2}$ bound state of the n$^{16}$O cluster system, which does not contain the FS, were constructed on the basis of the simple assumptions about the consistency of the scattering potential with the scattering phases, and about the consistency of the BS potential with the main characteristics of the GS of $^{17}$O (the binding energy, the charge radius, and the AC).

The results of calculation of the cross sections for the *E*1 transitions from the $^2P_{3/2}$ and $^2P_{1/2}$ scattering waves to the first excited state of $^{17}$O (process 2) are shown in Fig. 9b by the solid curve. This calculation was performed in the energy range from 10 keV to 1.3 MeV without the second resonance in the $^2P_{3/2}$ wave at 1312 keV,[94] i.e., for the $^2P_{1/2} + ^2P_{3/2} \to ^2S_{1/2}$ process. Here the dotted curve shows the results of measurements of the cross sections from Ref. 95 and 96 for the transition to the $1/2^+$ first excited level in the energy range 20–280 keV. The squares show the measurements of the total cross sections.[115] It seems that the measurement of Ref. 115 are in better agreement with the earlier experimental results for the transitions to the first excited state.[95,96] In this case, they are quite reasonably described in our calculations at energies 20–560 keV. For the potential of the non-resonance $^2P_{1/2}$ wave without the FS the zero depth was used, and for the potential of the resonance $^2P_{3/2}$ wave interaction of Eq. (49) was used.

By analogy with the earlier considered n$^{12}$C system,[89] one can assume that the second excited state with $J^\pi = 1/2^-$ at 3.055 MeV relative to the GS[94] may belong to the $^2P_{1/2}$ wave. Then one should accept the presence of the bound AS in this partial wave which as usual must lead to a zero scattering phase shift. In this case, for example, the parameters of the $^2P_{1/2}$ potential are accepted a little deeper than they were determined for the interaction with $J^\pi = 3/2^-$, (see of Eq. (49))

$$V_{1/2} = 1593.435350 \text{ MeV, and } \alpha_S = 6.0 \text{ fm}^{-2}. \qquad (52)$$

It gives the approximate to zero phase shifts in the region up to 1 MeV, and leads to the binding energy of -1.08824 MeV relative to the n$^{16}$O threshold at the accuracy of the FDM of $10^{-5}$ MeV. The charge radius of $^{17}$O in the $1/2^-$ second excited state is equal to 2.70 fm, the mass radius of 2.65 fm, and the AC is equal to 0.22 in at the interval of 2–18 fm. The results obtained for the total cross sections of the capture for the *E*1-process $^2P_{1/2} + ^2P_{3/2} \to ^2S_{1/2}$ with such potential and potential of Eq. (49) almost coincide with the previous ones; they are shown in Fig. 9b by the dashed curve.

It is not succeed to find the values of the AC for the second excited state at 3.055 MeV with $J^\pi = 1/2^-$, so it is not possible to compare the obtained above value of the AC. In order to, at least partially, get rid of the existing ambiguity of the parameters of this scattering potential let's consider another version of it with a wider interaction well

$$V_{1/2} = 270.711230 \text{ MeV, and } \alpha_S = 1.0 \text{ fm}^{-2}. \quad (53)$$

This potential also leads to a near-zero phase shifts, the binding energy of -1.08824 MeV with the accuracy of the FDM $10^{-5}$ MeV, the charge radius of 2.79 fm, the mass radius of 2.69 fm and the AC equal to 0.39 in the interval of 3–23 fm, which is nearly twice as much as for the previous potential. The capture total cross sections are shown in Fig. 9b by the dotted curve.

As is seen from these results the cross section does not strongly depend on the presence of the bound allowed state in the 1/2⁻ wave, if the width of such potential is comparable with the width of interaction in the $^2P_{3/2}$ wave and is in the region 1.0–6.0 fm, and the values of the AC are in the interval from 0.2 to 0.4. Hence, one cannot use this transition for the unambiguous choice of the form of the interaction potential in the $^2P_{1/2}$ scattering wave and determine the presence of the allowed bound state at $J = 1/2^-$ in it.

The total cross section for all transitions to the $5/2^+$ ground and the $1/2^+$ first excited states are shown in Fig. 9c. It is evident that the results of the calculations quite reproduce the data,[95,96] and the measurement results[115] are mainly below the calculated curve. Note that here the possible at the lowest energies $M1$ transitions have not been considered yet, but these transitions can increase somewhat the total cross section at the minimum energy shown in Fig. 9c.

In this way, the considered above methods of constructing of the clusters interaction potentials allow, on the whole, to reproduce correctly experimental data for the total cross sections of the radiative capture at energies from 20 keV to 560 keV. However, in the case of the $M1$ transition to the first excited state in this cluster system the used above criteria are not sufficient for the unambiguous definition of the potential, and, as will be seen below, it is necessary to vary its parameters to describe better the experimental data[78,113,114] at the lowest energy 25 meV.

Going to the low-energy region, note that below 100 eV the capture cross section is completely determined by the $M1$ process with the transition from the $^2S_{1/2}$ scattering wave to the first excited state of $^{17}O$ nucleus (process 4 in the given above list of transitions). The potential of the $^2S_{1/2}$ wave of the n$^{16}O$ scattering contains the forbidden state, as follows from the considered above classification of cluster states. This potential at the energies under consideration must lead to a zero phase shift, but as it has the forbidden state, its depth cannot be equal to zero. The form of this potential was refined in order to describe correctly the cross section at 25 meV. The parameters of this potential are

$$V_S = 10.0 \text{ MeV, and } \alpha_S = 0.03 \text{ fm}^{-2}. \quad (54)$$

The results of calculations of the total cross sections with the $M1$ transition at 10 meV–1.0 MeV are shown in Fig. 9d by the solid curve. At the energy 25 meV the triangles show the results of experimental measurements,[78,113,114] which are in the range 150–200 μb. The dashed curve in Fig. 9d corresponds to the results for the cross sections caused only by the $M1$ process, and the $E1$ transitions are shown in this figure by the dotted curve. As is seen from this figure the *E*1 cross section drops dramatically and at 100 eV is about three times smaller than the cross section of the *M*1 transition.

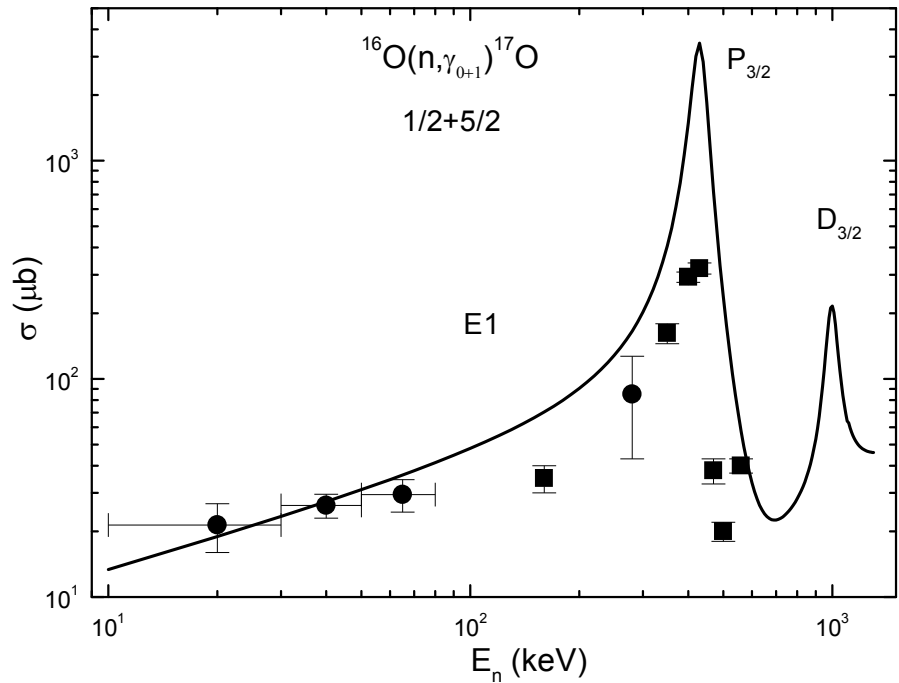

Fig. 9c. The total cross sections of the *E*1 radiative neutron capture on $^{16}$O to the $5/2^+$ ground and the $1/2^+$ first excited state of $^{17}$O. Experimental data: ● – Refs. 95 and 96, ■ – Ref. 115. The continuous curve – our calculation of the total cross section.

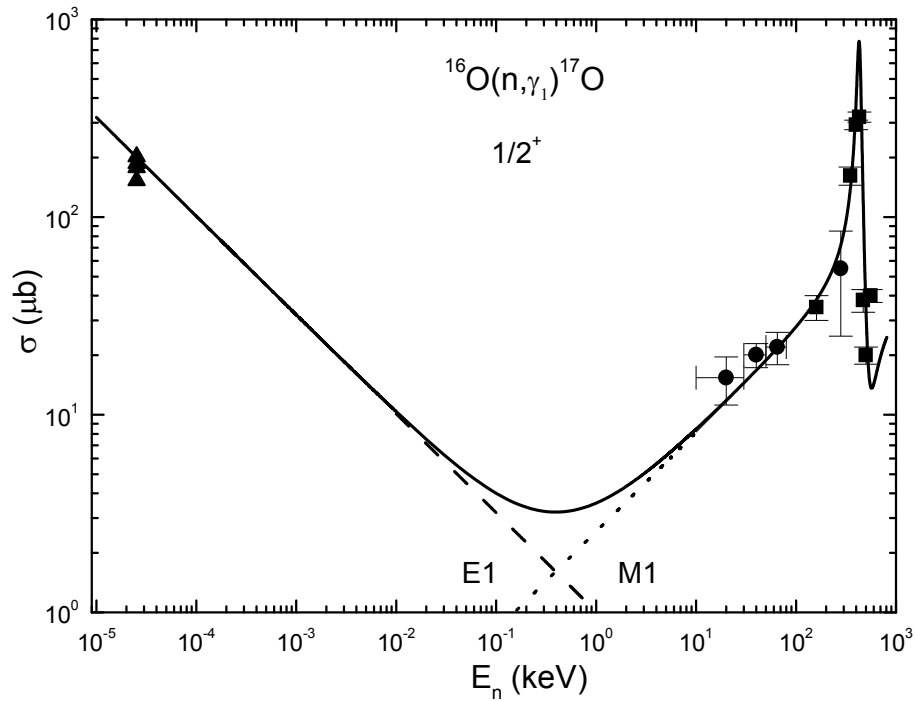

Fig. 9d. The total cross sections of E1 and M1 neutron radiative capture on 16O on the first excited 1/2+ state of 17O. Experimental data: ● – Refs. 95 and 96, ■ – Ref. 115, ▲ – Ref. 78, 113, and 114. The solid curve is our calculation of the total cross section for the potentials given in the text.

An additional point to emphasize is that only $^2S_{1/2}$ scattering potential with the FS allows one to describe correctly the capture total cross sections at the energy of 25 meV. If the potential without the FS is used, none of its parameters cannot reproduce correctly the behavior of the total cross sections at this energy. Namely, at the energy 25 MeV for potential of Eq. (54) $^2S_{1/2}$ scattering phase is equal to 0.00812° (in case of using the newer value of constant $\hbar^2/m_0 = 41.8016$ MeV·fm$^2$ the phase is equal to 0.00692°) and capture cross section takes a value 202 μb. In case of using of the $^2S_{1/2}$ potential without FS and with the parameter $V_0 = 0$ MeV, its phase shift equals zero, and the cross section at 25 meV increases up to 12.5 mb. Another variant of the $^2S_{1/2}$ potential without FS, for example, with the parameters $V_0 = 3.18$ MeV and $\alpha = 0.1$ fm$^{-2}$, gives the same scattering phase 0.00812° and leads to further increasing of the cross section, i.e. to the value 37.9 mb. Thereby, the description of the cross section at the energy 25 meV within the MPCM methods is possible only by using the $^2S_{1/2}$ scattering potential containing the forbidden state.

Such a great difference in the results may be explained by behavior of the scattering wave function of these potentials at 25 meV, as shown in Fig. 10. For the potential without FS and with $V_0 = 0$ at 25 meV it is the straight line (the dashed curve), with a depth -3.18 MeV it is the dotted line, and for interaction of Eq. (54) it oscillates even at such a low energy and has a node at 5.17 fm (the solid curve in Fig. 10).

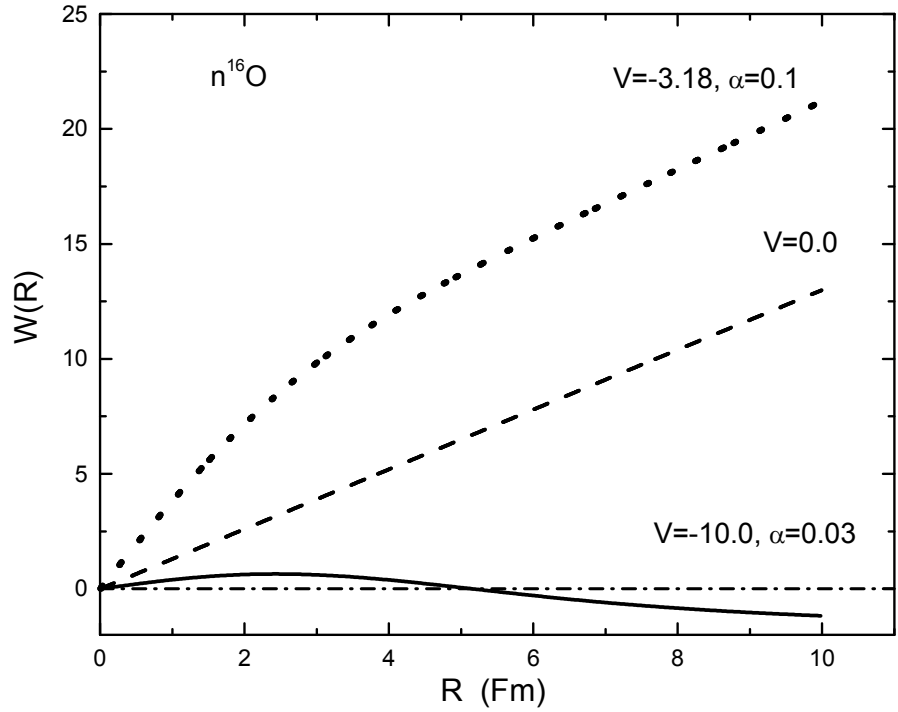

Fig. 10. Wave functions of the scattering potentials at 25 meV

Since at the energies from 10 meV to, approximately, 10 eV the calculated cross section shown in Fig. 9d by the solid curve is practically a straight line, it may be approximated by a simple function of the form (22). The value of the reduced constant $A = 1.0362$ μb·keV$^{1/2}$ were determined by one point in the cross sections at the minimum energy equal to $10^{-5}$ keV.

One may consider the module of the relative departure of the calculated theoretical cross section and the approximation of the cross-section (23) by this function in the range from 10 meV to 10 eV. At the energies below 10 eV, the relative departure is equal to 2.5%, and up to 1 eV does not exceed 0.5%. It may be assumed that this form of energy dependence of the total cross section will be stored at lower energies, and the cross section, for example, at the energy 1 μeV, takes the value 32.8 mb.

## 7. Conclusion

Thus, it is possible to acceptably describe the available experimental data on the total cross-sections of neutron capture on $^9$Be at the energies at 25 meV and 25 keV if certain assumptions are made concerning the general character relative to the interaction potentials in the n$^9$Be channel of $^{10}$Be.

In the case of consideration of the $E$1 transition from the $^3D_3$ resonance scattering wave at the energy of 622 keV to the first and second excited states of $^{10}$Be in the capture cross-sections, the observed resonance reaches about 2.6 mb. The usage of the potentials for the $^3D_3$ resonance phase shift with different number of FSs has almost no effect on the resonance value, which influences the cross-sections at energies above 0.2 MeV. Taking into account the transitions to the fifth ES and the resonance at 812 keV does not lead to the essential changes of the summed cross section shown in Fig. 2b. Therefore, it always can be considered that the $^3D$ waves contain one bound FS with tableau {541} and the state with {442} is unbound. All $^3P$ potentials contain one bound FS with tableau {541}, and $^3P_0$ wave includes another allowed BS, corresponding to the GS of $^{10}$Be. At the same time the $^3S_1$ wave contains one allowed BS with tableau {442}, and the FS are absent. Thus, all considered potentials quite correspond to the given above classification of FSs and ASs according to Young tableaux, and allow one to correctly describe available experimental data under consideration of few possible $E$1 transitions between different levels of $^{10}$Be.

We should note that the experimental data on the cross-sections of the considered reaction that are available for our disposal are obviously insufficient. Furthermore, more thorough consideration of the $n^9$Be → $^{10}$Beγ radiative capture at thermal and astrophysical energies is apparently needed, especially in the range of the resonance with $J^\pi$ = 3$^-$ at 0.622 MeV.

The possibility of description of the experimental data on total cross sections of the radiative neutron capture on $^{14}$C and $^{14}$N is considered within the frame of the modified potential cluster model with forbidden states and their classification according to Young tableaux. It is shown that the using model and the potential construction methods allow us to reproduce correctly the behavior of experimental cross sections at the energies from 10 meV (10$^{-2}$ eV) to 1 MeV. Thereby, it is possible to coordinate description of the elastic scattering processes (phase shifts), main BS characteristics of nuclei (energy, radius, AC) and total radiative capture cross sections for all considered above processes of radiative capture on the basis of unified variants of intercluster potentials for each reaction. In addition, all these results are obtained on the basis of the general classification of orbital cluster states according to Young tableaux.[6,7,9]

The possibility of description of experimental data for the total cross-sections of radiative neutron capture on $^{15}$N at energies from 25 to 370 keV was considered within the framework of the MPCM with the classification orbital states according to Young tableaux. It was shown that it is well succeeded to explain the value of the total cross section in the considered energy range only on the basis of the $E$1 transitions from the different states of the n$^{15}$N scattering to the ground and excited states of $^{16}$N in the n$^{15}$N channel.

In this way, the quite acceptable results for description of the experimental total cross sections of the radiative neutron capture on $^{16}$O at the energies from 25 meV to 560 keV have been obtained for the considered intercluster potentials of the n$^{16}$O interaction, which answer to the classification of the states according to Young tableaux. The potential of the $^2S_{1/2}$ scattering wave with the FS, the phase of which is close to zero, have been obtained. This potential allows us to describe correctly the behavior of the experimental cross sections at the lowest energies. It was shown that the description of the low-energy cross sections is possible only if there is a forbidden state in such potential. Interaction of the $^2S_{1/2}$ scattering wave without the FS, in principle, does not allow us to reproduce correctly the value of the capture cross section at such a low energy.

In the future it is necessary to carry out new experimental measurements of the total capture cross sections at 1 eV–1 keV, where the calculations predict the well-defined behavior of the cross-sections with a smooth minimum at the energy 0.4 keV and with the value 3 μb (Fig. 9d). In addition, near 1.0 MeV, i.e., in the region of the $^2D_{3/2}$ resonance, a well-defined value of the second maximum of the cross sections (Fig. 9a) is also received. In both cases obtained results are somewhat different from the similar data obtained earlier in Ref. 112 and, apparently, only the new experimental measurements are able to eliminate this discrepancy.

Table 1. The characteristics of nuclei and cluster systems, and references to works in which they were considered.[+]

| No. | Nucleus ($J^\pi$,$T$) | Cluster channel | $T_z$ | $T$ | Refs. |
|---|---|---|---|---|---|
| 1. | $^3$H (1/2$^+$,1/2) | n$^2$H | -1/2 + 0 = -1/2 | 1/2 | 6, 9, 33 |
| 2. | $^3$He (1/2$^+$,1/2) | p$^2$H | +1/2 + 0 = +1/2 | 1/2 | 7, 33, 87 |
| 3. | $^4$He (0$^+$,0) | p$^3$H | +1/2 - 1/2 = 0 | 0 + 1 | 6, 9, 10 |
| 4. | $^6$Li (1$^+$,0) | $^2$H$^4$He | 0 + 0 = 0 | 0 | 6, 9, 10 |
| 5. | $^7$Li (3/2$^-$,1/2) | $^3$H$^4$He | -1/2 + 0 = -1/2 | 1/2 | 6, 9, 10 |
| 6. | $^7$Be (3/2$^-$,1/2) | $^3$He$^4$He | +1/2 + 0 = +1/2 | 1/2 | 6, 9, 10 |
| 7. | $^7$Be (3/2$^-$,1/2) | p$^6$Li | +1/2 + 0 = +1/2 | 1/2 | 6, 11, 116 |
| 8. | $^7$Li (3/2$^-$,1/2) | n$^6$Li | -1/2 + 0 = -1/2 | 1/2 | 7, 89, 117 |
| 9. | $^8$Be (0$^+$,0) | p$^7$Li | +1/2 - 1/2 = 0 | 0 + 1 | 6, 10, 11 |
| 10. | $^8$Li (2$^+$,1) | n$^7$Li | -1/2 - 1/2 = -1 | 1 | 7, 43, 118 |
| 11. | $^{10}$B (3$^+$,0) | p$^9$Be | +1/2 - 1/2 = 0 | 0 + 1 | 6, 9, 10 |
| 12. | $^{10}$Be (0$^+$,1) | n$^9$Be | -1/2 - 1/2 = -1 | 1 | 121, Present work |
| 13. | $^{13}$N (1/2$^-$,1/2) | p$^{12}$C | +1/2 + 0 = +1/2 | 1/2 | 6, 11 |
| 14. | $^{13}$C (1/2$^-$,1/2) | n$^{12}$C | -1/2 + 0 = -1/2 | 1/2 | 7, 44, 89 |
| 15. | $^{14}$N (1$^+$,0) | p$^{13}$C | +1/2 - 1/2 = 0 | 0 + 1 | 63, 88 |
| 16. | $^{14}$C (0$^+$,1) | n$^{13}$C | -1/2 - 1/2 = -1 | 1 | 7, 44, 88 |
| 17. | $^{15}$C (1/2$^+$,3/2) | n$^{14}$C | -1/2 – 1 = -3/2 | 3/2 | 7, 118, Present work |
| 18. | $^{15}$N (1/2$^-$,1/2) | n$^{14}$N | -1/2 + 0 = -1/2 | 1/2 | 7, 118, Present work |
| 19. | $^{16}$N (2$^-$,1) | n$^{15}$N | -1/2 - 1/2 = -1 | 1 | 7, 119, Present work |
| 20. | $^{16}$O (0$^+$,0) | $^4$He$^{12}$C | 0 + 0 = 0 | 0 | 10, 120 |
| 21. | $^{17}$O (5/2$^+$,1/2) | n$^{16}$O | -1/2 + 0 = -1/2 | 1/2 | 122, Present work |

[+] $T$ – isospin and $T_z$ – its projection, $J^\pi$ – total moment and parity.

In conclusion note that there are twenty one cluster system considered on the basis of the modified potential cluster model with the classification of the orbital states according to the Young tableaux, in which it is possible to obtain acceptable results on the description of the characteristics of the processes of the radiative capture of nucleons or light clusters, mainly, on the 1$p$ shell nuclei. The properties of these cluster nuclei, some characteristics and the considered cluster channels are shown in Table 1. Recent results are presented in Refs. 43, 44, 63, 87, 89, 116, 117, 118, 119, 120, 121, and 122.

## Acknowledgments

The work was performed under the grant No. 0151/GF2 "Studying of the thermonuclear processes in the primordial nucleosynthesis of the Universe" of the Ministry of Education and Science of the Republic of Kazakhstan.

In conclusion, the authors express their deep gratitude to Blokhintsev L.D., Burkova N.A., Mukhamedzhanov A., Yarmukhamedov R. for extremely useful discussion of the certain parts of this work.